\documentclass[journal=jacsat, manuscript=article]{achemso}

\usepackage[version=3]{mhchem}
\usepackage{multicol}
\usepackage[utf8]{inputenc}
\usepackage{graphicx}
\usepackage{wasysym}
\usepackage{natbib}
\usepackage{amsmath}
\usepackage{amsfonts}
\usepackage{amssymb}
\usepackage{xcolor}
\usepackage{soul}
\usepackage{setspace}
\usepackage{dcolumn}
\usepackage{color}
\usepackage{bm}
\usepackage{graphicx,wrapfig,lipsum}
\usepackage[normalem]{ulem}
\usepackage{pdfpages}

\usepackage[labelfont=bf]{caption}

\usepackage{url}

\usepackage{url}
\usepackage[ colorlinks = true,
linkcolor = red,
urlcolor  = blue,
citecolor = blue,
anchorcolor = red]{hyperref}
\usepackage{bibentry}
\usepackage{fancyhdr}
\author{Ranit Dutta}
\affiliation{These authors contributed equally to this work}
\alsoaffiliation[University]{Department of Physics, Indian Institute of Science, Bangalore 560012, India}

\author{Ayan Ghosh}
\affiliation{These authors contributed equally to this work}
\alsoaffiliation[University]{Department of Physics, Indian Institute of Science, Bangalore 560012, India}

\author{Kenji Watanabe}
\affiliation[University]{Research Center for Functional Materials, National Institute for Materials Science, 1-1 Namiki, Tsukuba 305-0044, Japan}
\author{Takashi Taniguchi}
\affiliation[University]{International Center for Material Nanoarchitectonics, National Institute for Materials Science,  1-1 Namiki, Tsukuba 305-0044, Japan}

\author{Anindya Das}
\email{anindya@iisc.ac.in}
\affiliation[University]{Department of Physics, Indian Institute of Science, Bangalore 560012, India}

\title{Resistance hysteresis in twisted bilayer graphene: Intrinsic versus extrinsic effects}

\begin{document}

 \begin{abstract}
    \begin{spacing}{2}

{\bf Hysteresis in resistance under magnetic field sweeps is a key signature for identifying magnetism in twisted bilayer graphene and similar systems. However, such sweeps can induce extrinsic thermal effects, complicating interpretations. Distinguishing intrinsic magnetic ordering from extrinsic thermal influences is crucial. In this study, we report hysteresis in the longitudinal resistance (\(R_{xx}\)) of a near magic-angle twisted bilayer graphene (TBG) sample under an in-plane magnetic field (\(B_{||}\)). The hysteresis phase appears at the edge of the superconducting dome, diminishes deep within the superconducting regime, and reemerges near the superconducting critical temperature (\(T \sim T_c\)). The hysteresis magnitude and coercive fields strongly depend on the magnetic field sweep rate (\(dB/dt\)) and exhibit transient relaxation in time-series measurements. Notably, similar hysteresis behavior was observed in the temperature profile of the sample stage, measured using a calibrated temperature sensor under analogous magnetic field cycles, suggesting extrinsic thermal origins rather than intrinsic magnetic ordering. These findings underscore the importance of carefully distinguishing intrinsic and extrinsic effects in resistance hysteresis observed in mesoscopic van der Waals systems.} 

\end{spacing}

    \end{abstract}


\begin{spacing}{2}
 Breaking of the sublattice symmetry in two-dimensional materials such as monolayer graphene leads to the formation of time-reversal symmetric Chern bands at K and K$'$ valleys with equal and opposite non-zero Chern numbers by virtue of a finite mass acquired by the Dirac bands \cite{song2015topological}. Twisted moir\'e structures such as twisted bilayer graphene (TBG) are a platform to engineer and host bands with non-zero Chern numbers that can be influenced by strong electron-electron interaction effects when the bands become extremely flat \cite{bistritzer2011moire,cao2018unconventional1,cao2018correlated1,lu2019superconductors}. Breaking of the time-reversal symmetry of such Chern bands under strong interactions \cite{liu2021theories} can lead to the formation of Chern insulators with the signature of quantized anomalous Hall effect (QAH) \cite{hejazi2021hybrid,kwan2021exciton} and emergent ferromagnetism with anomalous Hall effect \cite{repellin2020ferromagnetism,chatterjee2020symmetry,bultinck2020mechanism,wu2020collective,zhang2019twisted,shi2021moire}. In experiments, anomalous Hall effect (AHE)/ferromagnetism and QAH states have already been realized at moir\'e filling factors of $\nu=1$, $2$, $3$ in magic-angle twisted bilayer graphene (MATBG)  \cite{stepanov2021competing,lin2022spin,sharpe2019emergent}, twisted bilayer graphene aligned to hBN \cite{sharpe2021evidence}, 
 ABC-trilayer graphene aligned to hBN \cite{chen2022tunable}, twisted double bilayer graphene (TDBLG) proximitized to WSe$_2$ \cite{kuiri2022spontaneous}, twisted monolayer bilayer graphene \cite{polshyn2020electrical} and many more.

 Owing to the vanishing spin-orbit coupling of graphene, the ferromagnetic order of spin origin is not expected in the superlattice structures formed exclusively from graphene layers.
 As a result, the orbital component becomes the driving force behind the ferromagnetic behavior observed in these moir\'e platforms and is identified as `orbital magnetism' \cite{lu2019superconductors,polshyn2020electrical}. The AHE and orbital 
 ordering are identified through the hysteresis observed in the 
 transverse and (/or) longitudinal magnetoresistance under an applied perpendicular magnetic field. The hysteresis in resistance ($R_{xy}/R_{xx}$) is a direct consequence of the hysteresis of the magnetization of the
 (Ferro) magnetic ordering under a magnetic field, similar to what is found in traditional ferromagnets. Recently, such hysteresis curves have been reported in other graphene-based systems like suspended rhombohedral trilayer graphene (s-RTG) \cite{lee2022gate} and a superconducting magic-angle twisted trilayer graphene (MATLG) \cite{mukherjee2024superconducting} under the application of parallel magnetic field ($B_{||}$), suggesting existence of magnetic ordering of spin nature in these van der Waals platforms. 
 
The hysteresis in $R_{xy}$ (and$/$or $R_{xx}$) magnetoresistance is characterized by: i) sharp/step jumps in measured resistance at the coercive field ($|B_{co}|$), which depends on the direction of the magnetic field sweep and properties of the sample,  and ii) a mismatch in the residual resistance at zero magnetic fields under a complete magnetization loop. For mesoscopic structures such as twisted moir\'e heterostructures, which are routinely measured under sub-Kelvin temperatures, it becomes crucial to differentiate between the hysteresis in resistance occurring due to intrinsic magnetic ordering and the possible impact of change in resistance due to a rise and fall of the effective temperature of the sample itself facilitated extrinsically by the continuous sweep of the magnetic field, 
intrinsic to the sample holder, influencing the experimental results. Thus, observation of hysteresis in resistance to the magnetic field in `twisted' mesoscopic devices requires careful deliberation to uncover the exact intrinsic or extrinsic nature to infer any meaningful conclusion.

In this letter, we present 
an experimental study on the observation of hysteresis in longitudinal resistance ($R_{xx}$) under an in-plane magnetic field ($B_{||}$) in a near magic-angle TBG ($\theta=0.95^\circ$) that hosts a superconducting (SC) phase when it is electron-doped to a moir\'e filling factor of $\nu\sim2-3$ \cite{dutta2025electric}. This hysteresis phase is largely prominent around the SC phase, with $R_{xx}$ hysteresis vanishing deep inside the superconducting region. Additionally, the hysteresis phase is recovered once superconductivity is diminished, as seen with increased temperature around superconducting critical temperature, $T_c$ \cite{dutta2025electric}. We have found that the magnitude of reduced magnetoresistance (rMR) and coercive fields strongly depends on how rapidly (ramp-rate, $dB/dt$) the magnetic field is cycled through a closed loop. Also, $R_{xx}$ is found to have a transient relaxation profile in a `time-series' experiment. Compared with an independently calibrated temperature sensor similarly mounted to the sample stage, we find a similar hysteresis behavior under the applied magnetic field in the temperature of the sample stage itself that matches qualitatively in an analogous fashion with that of the near magic-angle TBG. Our results suggest that the observed hysteresis in $R_{xx}(B_{||})$ could be due to the result of an extrinsic effect guided by the change in the temperature profile of the sample stage subjected to the continuous change of the magnetic field, and indicates a requirement for further studies in these layered van der Waals platforms.

 \begin{figure*}[htbp!]
	\centering
	\includegraphics[width=1.0\textwidth]{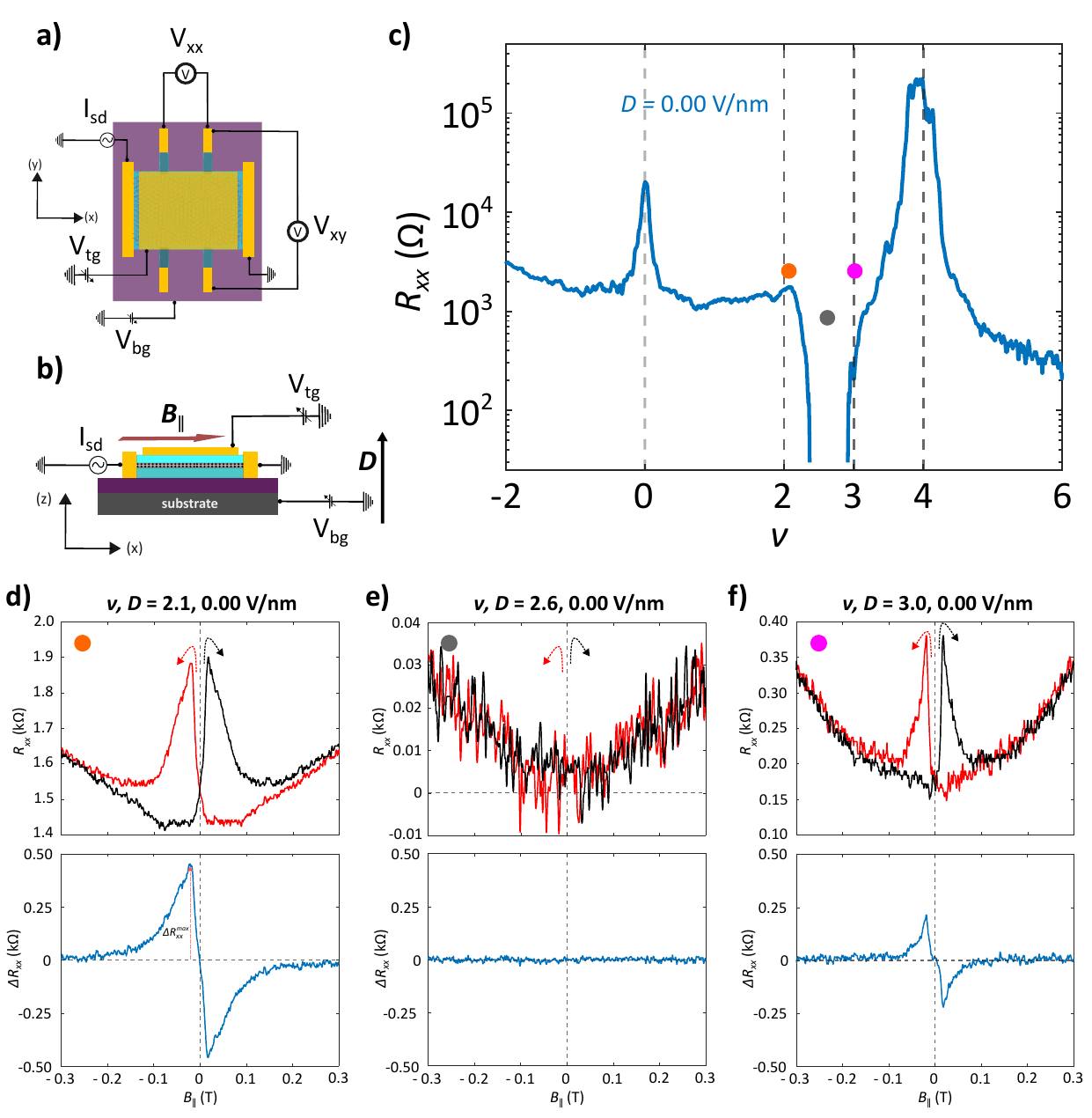}
	\caption[Measurement scheme, device and hysteresis response]{\textbf{Measurement scheme, device and hysteresis response:} \textbf{a)} Schematic `top view' of the Hall-bar geometry device with four probe measurements set up to measure $V_{xx}$ and $V_{xy}$. \textbf{b)}  $V_{bg}$ and $V_{tg}$ is applied simultaneously in the dual-gate geometry to tune $n$ and $D$. We apply an in-plane magnetic field, $B_{||}$, for our experiments.  \textbf{c)} $R_{xx}$ vs. filling factor, $\nu$, at zero $D$ for $-2 < \nu < 6$ at a base temperature of $20$ mK in absence of any magnetic field. The three solid circles correspond to filling ($\nu$) values: $2.1$ (orange solid circle), $2.6$ (gray solid circle), and $3.0$ (magenta solid circle)
  respectively at $D=0.00$ V/nm. Magnetoresistance curve measured for zero $D$ at \textbf{d)} $\nu=2.1$, \textbf{e)} $\nu=2.6$, and \textbf{f)} $\nu=3.0$. We have defined the red $R_{xx}$ trace with $B_{||}$ $(+300 $ mT $\rightarrow-300$ mT) as $R^{FW}_{xx}$. Similarly, the black $R_{xx}$ trace with  $B_{||}$ $(-300 $ mT $\rightarrow+300$ mT) sweep has been defined as $R^{BW}_{xx}$. Calculated $\Delta R_{xx}$ (bottom panel) within the $B_{||}$  sweep range of $\pm300$ mT. The $B_{||}$ sweep rate is: $100$ mT/min. The average operational temperature of $B_{||}$ sweep was $\sim70$ mK.}
	\label{fig:1}
\end{figure*}



For our study, we have fabricated a dual-gated hBN encapsulated twisted bilayer graphene device using the modified `cut and stack' \cite{paul2022interaction,ghosh2023evidence,dutta2025electric} technique from an exfoliated single monolayer graphene flake on a SiO$_2/$Si substrate. 
The dual-gated device structure helps to tune the number density, $n$, and the perpendicular displacement field, $D$, independently (see \textcolor{magenta}{Supplementary Information SI- 8} ) in our TBG device. Figure. \ref{fig:1}a shows a schematic of the measurement scheme to measure longitudinal ($R_{xx}$) and transverse ($R_{xy}$) resistance in a standard Hall-bar geometry setup. We measured low-temperature transport in a cryo-free dilution refrigerator at $\sim20-25$ mK. 
The four-probe longitudinal ($V_{xx}$) and transverse voltage ($V_{xy}$) were measured using a lock-in amplifier at a low frequency ($\sim13$ Hz). Measured $R_{xx}$ in Figure. \ref{fig:1}c shows the appearance of the superconducting phase between filling factor, $\nu (\equiv 4n/n_s)\approx 2-3$. We have applied a parallel magnetic field, $B_{||}$ (see Figure. \ref{fig:1}b and \textcolor{magenta}{SI- 1}),  
to study the hysteresis behavior in resistance with magnetic field sweep direction in presence of external tuning knobs like $n$, $T$ in our TBG device. 

 We have measured $R_{xx}$ at different $\nu$ values at zero $D$ by sweeping the in-plane magnetic field, $B_{||}$, within the range of $\pm 300$ mT. For the hysteresis experiment for a given $\nu$; $B_{||}$ is swept continuously from $+300$ mT to $-300$ mT, which we define as the `forward (FW)' sweep and then back again from $-300$ mT to $+300$ mT defined as the `backward (BW)' sweep. The reduced magnetoresistance (rMR) is defined as:

 \begin{equation}
	\label{BMR}
	\Delta R_{xx} = R^{FW}_{xx} (\textcolor{red}{+B_{||}\rightarrow-B_{||}})-R^{BW}_{xx} (\textcolor{black}{-B_{||}\rightarrow+B_{||}})
\end{equation}

\begin{figure*}[htbp!]
	\centering
	\includegraphics[width=1.0\textwidth]{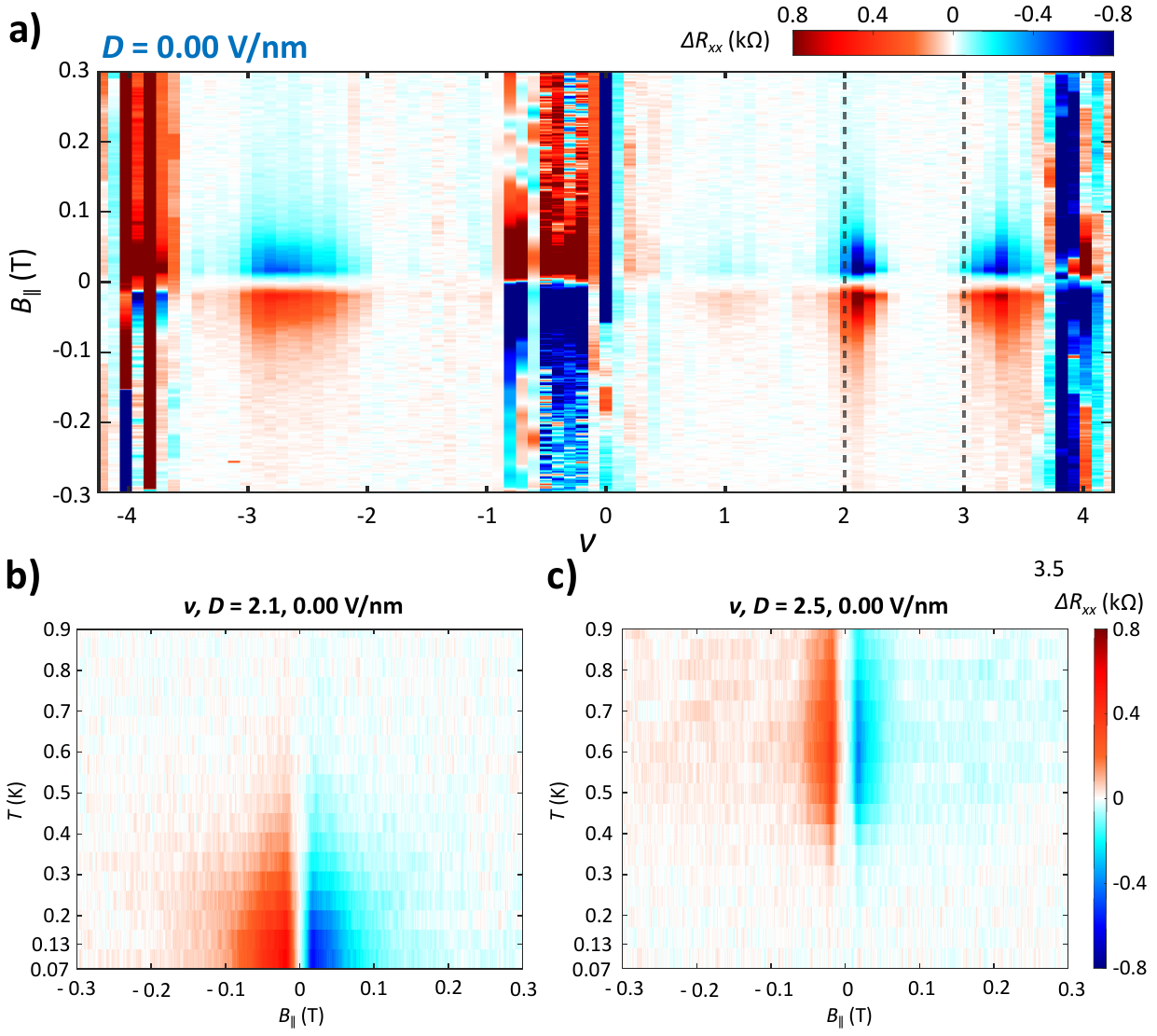}
	\caption[rMR map in the $\nu-B_{||}$ phase and evolution of the hysteresis phase with $T$]{\textbf{rMR map in the $\nu-B_{||}$ phase and evolution of the hysteresis phase with $T$:} \textbf{a)} $\Delta R_{xx} (\nu,B_{||})$ $2-$d colormap for $D=0.00$ V/nm. $\Delta R_{xx} (B_{||},T)$ $2-$d colormap at $D=0.00$ V/nm for \textbf{b)} $\nu=2.1$ and \textbf{c)} $\nu=2.5$.}    
	\label{fig:2}
\end{figure*}

Near the optimal doping of $\nu_{op}=2.7$ 
of the SC phase of the near magic-angle TBG, the forward and backward $R_{xx}$ traces are almost identical, as shown in Figure. \ref{fig:1}e (top panel) with no signature of hysteresis behavior in resistance with $B_{||}$. From the definition of rMR defined earlier, we then get $\Delta R_{xx}\approx0$ (Figure. \ref{fig:1}e (bottom panel)). But near the underdoping ($\nu<\nu_{op}$) and the overdoping ($\nu>\nu_{op}$) region, we observe a hysteresis between the measured $R_{xx}$ for the two different sweep directions 
of $B_{||}$ resembling the shape of the wings of a `butterfly' as shown in Figure \ref{fig:1}d,f (top panel). Curiously, the sharp peaks on both sides of the $B_{||}=0$ appears at $|B_{||}|\sim17-19$ mT. As demonstrated in Figure. \ref{fig:1}d,f (bottom panel), depending on the structure of the Butterfly Magnetoresistance (BMR) \cite{ohta2021butterfly,taniguchi2020butterfly,mukherjee2010anomalous,li2009origin} hysteresis curves (\textcolor{magenta}{SI- 2}) in Figure. \ref{fig:1}d,f (top panel), we get a positive (negative) value of $\Delta R_{xx}$ for a negative (positive) $B_{||}$ following equation (\ref{BMR}). These observations of hysteresis behavior in the magnetoresistance around the superconducting region can be seen more easily from the $\Delta R_{xx} (\nu, B_{||})$ $2-$d colormap in Figure. \ref{fig:2}a  for $D=0.00$ V/nm. At zero $D$, we see two separate hysteresis `islands' around $\nu\sim2.2$ and $\nu\sim3$ flanking either side of the SC region of the near magic-angle TBG. 


 Figure. \ref{fig:2}b,c show the $\Delta R_{xx} (B_{||},T)$ $2-$d colormap at zero $D$ for two filling factors, one near the `underdoping' of the SC phase at $\nu=2.1$, and one
 inside the SC phase, $\nu=2.5$. From Figure. \ref{fig:2}a, we have seen the presence and absence, respectively, of distinct hysteresis signatures around these $\nu$ values at the operational sample temperature of $\sim 70 $mK. Still, as soon as we increase the effective sample (/bath) temperature, we observe strikingly opposite behaviors for these two fillings. For $\nu=2.1$ in Figure. \ref{fig:2}b, which sits at the left `island', $\Delta R_{xx}$ gradually diminishes and vanishes completely ($\Delta R_{xx}\approx0$) for $T > 650$ mK. On the other hand, for $\nu=2.5$ in Figure. \ref{fig:2}c, where we don't see any sign of hysteresis, a hysteresis phase emerges
 with increasing $T$ above $\sim400$ mK and gets stronger at $T\sim800$ mK around the $T_c$ of the SC phase \cite{dutta2025electric}. These observations highlight the emergence of the hysteresis phase with a gradual weakening of the SC phase, which is achieved here by increasing the sample (TBG) temperature around the $T_c$.



\begin{figure*}[htbp!]
	\centering
	\includegraphics[width=1.0\textwidth]{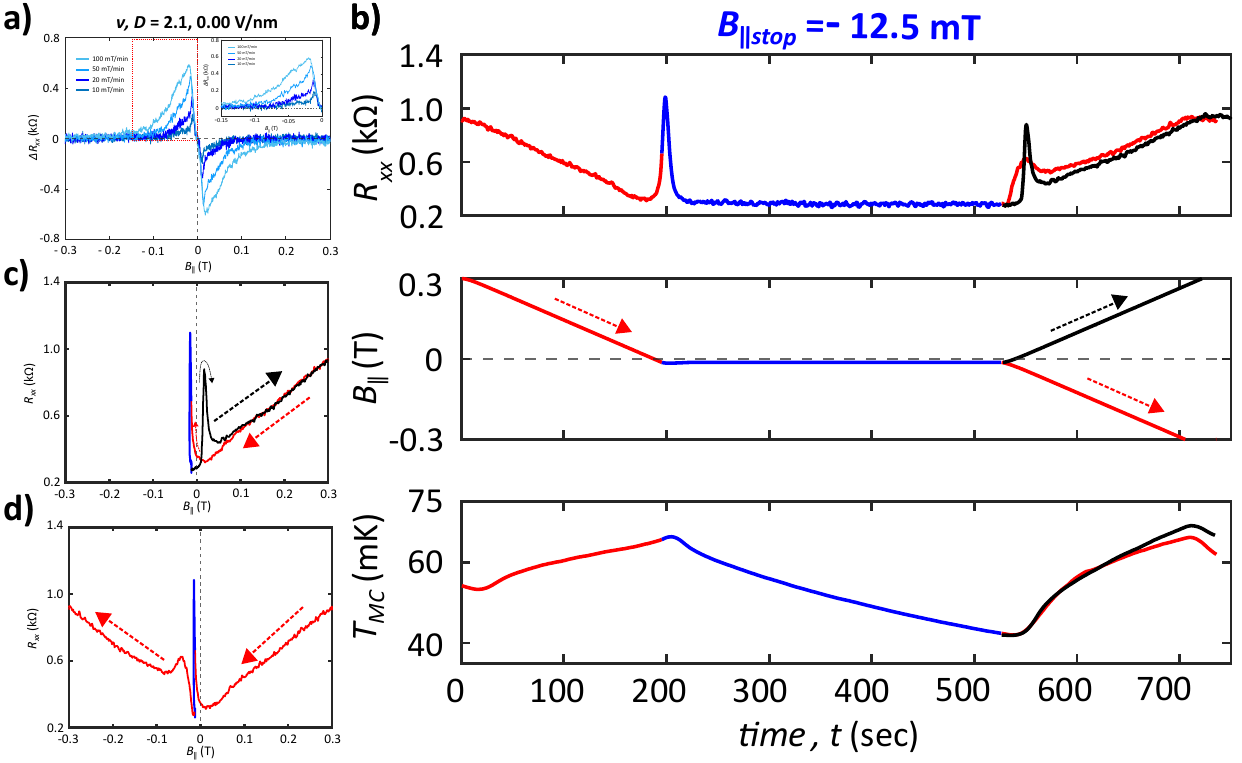}
	\caption[Effect of ramp-rate of $B_{||}$ and transient relaxation of $R_{xx}$]{\textbf{Effect of ramp-rate of $B_{||}$ and transient relaxation of $R_{xx}$:} \textbf{a)} $\Delta R_{xx}$ vs. $B_{||}$ at $(\nu,D)=2.1,0.00$ V/nm for four different ramp-rates of $B_{||}$ : $100$ mT/min, $50$ mT/min, $20$ mT/min, and $10$ mT/min. The inset shows a zoomed-in region highlighted in a red dashed box. \textbf{b)}  $R_{xx}$, $B_{||}$, and $T_{MC}$ with time, $t$, plotted in three panels for $B_{||stop} =-12.5$ mT. $R_{xx}$ transiently relaxes to a steady state value at the beginning of the `stop interval' of $300$ seconds. \textcolor{red}{Red} and Black arrows in the middle panel indicate the direction of the magnetic sweep performed before and after the `stop interval'. \textbf{c)} $R_{xx}$ vs. $B_{||}$ trace for $+300$ mT $\rightarrow B_{||stop} \rightarrow + 300$ mT.  \textbf{d)} $R_{xx}$ vs. $B_{||}$ trace for $+300$ mT $\rightarrow B_{||stop} \rightarrow -300$ mT.}
	\label{fig:3}
\end{figure*}

 Figure. \ref{fig:3}a shows $\Delta R_{xx}$ vs. $B_{||}$ at $(\nu,D)=2.1,0.00$ V/nm for four different ramp-rates of $B_{||}$: $100$ mT/min, $50$ mT/min, $20$ mT/min, and $10$ mT/min. The $|\Delta R_{xx}|$ 
 for $\pm B_{||}$ diminishes with a slower ramp-rate \cite{candini2011hysteresis}. Still, even for the slowest ramp-rate, we could get a significant magnitude of  $\Delta R_{xx}$ outlining the robustness of the hysteresis observed here. The anomalous behavior, though, is revealed through the shift in the positions of $B^{m}_{||}$ where we get the maxima and minima of $\Delta R_{xx}$ at different magnetic field ramp-rates for the magnetic sweep in the forward and backward direction. We find that the higher the ramp-rate, the higher is the $|B^{m}_{||}|$ highlighted in the inset of Figure. \ref{fig:3}a for $B_{||}<0$. 

Another surprising behavior is revealed for the observed hysteresis in $R_{xx} (B_{||})$ through the `time-series' experiment (for detailed measurement protocol, see \textcolor{magenta}{SI- 5}). In brief, for a given $\nu$ with a BMR hysteresis, we start sweeping the applied $B_{||}$ continuously from $+300$ mT to a field value of $B_{||stop} (\in [-300$ mT, $300$ mT$])$, the magnetic field is stopped sweeping and held at the $B_{||stop}$ for a `stop interval' of $300$ seconds and then the in-plane field is again swept from $B_{||stop}$ either to $+300$ mT or $-300$ mT. One such sweep is shown in Figure. \ref{fig:3}b for $B_{||stop}=-12.5$ mT, where $R_{xx}$, $B_{||}$ and temperature of the mixing chamber plate, $T_{MC}$ (\textcolor{magenta}{SI-Fig.  27b}), is plotted with time, $t$, in three vertical panels. Let us break down the behavior of $R_{xx}$ observed for $B_{||stop}=-12.5$ mT with $t$ in the top panel of Figure. \ref{fig:3}b. With the continuous sweep of $B_{||}$ from $+300$ mT to $-12.5$ mT, $R_{xx}$ traces the familiar trace of $R^{FW}_{xx}$ (red curve) as shown Figure. \ref{fig:1}d (top panel) but in time, $t$. We should note here that the value of $B_{||}$ where $R^{FW}_{xx}$ peaks with magnetic field sweep is $\sim-17.5$ mT, which signifies the coercive field value for the supposed switching of the magnetization direction of the magnetic ordering from one to another with the change in the direction of the applied field (\textcolor{magenta}{SI- 2}). 
Now, as soon as the field is stopped sweeping and is held at $-12.5$ mT for the next $300$ seconds, $R_{xx}$ rises sharply, forming a peak and subsequently relaxes transiently \cite{daptary2017observation} to a steady state value (blue curve) with time, $t$. This transient effect is quite surprising, and one may think this could be due to the settling of the moments of the TBG in the energetically favored magnetic ordered state at the $B_{||stop}$. When we start ramping the magnetic field again after the stop interval from $B_{||stop}$ (\textit{i.e.},$-12.5$ mT) towards $+300$ mT we see the $R_{xx}$ curve traces the familiar path of $R^{BW}_{xx}$ (black curve) as shown Figure. \ref{fig:1}d (top panel) with a $R_{xx}$ peak appearing with $t$ at a time when $B_{||}=+17$ mT.
The $R_{xx}$ vs. $B_{||}$ for magnetic field sweep performed as $+300$ mT $\rightarrow B_{||stop}(-12.5)$ mT $\rightarrow$ stop interval$\rightarrow+300$ mT is shown in Figure. \ref{fig:3}c. But what happens if we start sweeping the magnet from $B_{||stop}$ towards $-300$ mT after the stop interval? Strikingly in the time-space, we observe a 
broadened peak (second red curve in Figure. \ref{fig:3}b (top panel)) appearing as soon we start ramping the magnet before settling to the similar values of $R^{FW}_{xx}$ for $-300$ mT $<B_{||}<-100$ mT when compared with the red curve in Figure. \ref{fig:1}d (top panel). If the system is expected to have already settled in a favorable ordered state under the aegis of $B_{||stop}$ for a sweep direction of $+300$ mT $\rightarrow -300$ mT, the appearance of the peak again in a sweep towards $B_{||stop} \rightarrow -300$ mT indicates an anomaly. The transient effect and observing the peak while sweeping both the directions after $B_{||stop}$ raises a question about whether the observed hysteresis in $R_{xx}$ is related to intrinsic magnetic order or some external effects, 
which can give rise to such hysteresis behavior in TBG with applied $B$. 
Similar `time-series' behavior of other $B_{||stop}$ values are shown in \textcolor{magenta}{SI-Fig.  5,6,7}.
 
Even though we see that the change in the temperature of the MC plate while sweeping the magnet doesn't have any impact on the appearance or absence of BMR (\textcolor{magenta}{SI-Fig.  3}); during the transient experiment, we observe a sharp drop in the $R_{xx}$ to a steady state value as soon as $B_{||}$ is stopped sweeping and held at $B_{||stop}$. Similarly, we found that the temperature of the MC plate also decreased steadily during the `stop interval' from the continuous sweep of the magnetic field, which induced a higher temperature value, as shown in Figure. \ref{fig:3}b (bottom panel). This observation can raise a question of whether the continuous application of a magnetic field changes the local temperature of the sample stage at the end of the sample holder, which is situated at a substantial distance away from the MC plate stage of the cryo-free fridge (\textcolor{magenta}{SI-Fig.  27b}). If that is the case, then the measured $R_{xx}$ of the TBG could also show a change in resistance depending on the effective temperature experienced by the sample and the nature of $R(T)$ of the sample for a given $\nu$.

\begin{figure*}[htbp!]
	\centering
	\includegraphics[width=1.0\textwidth]{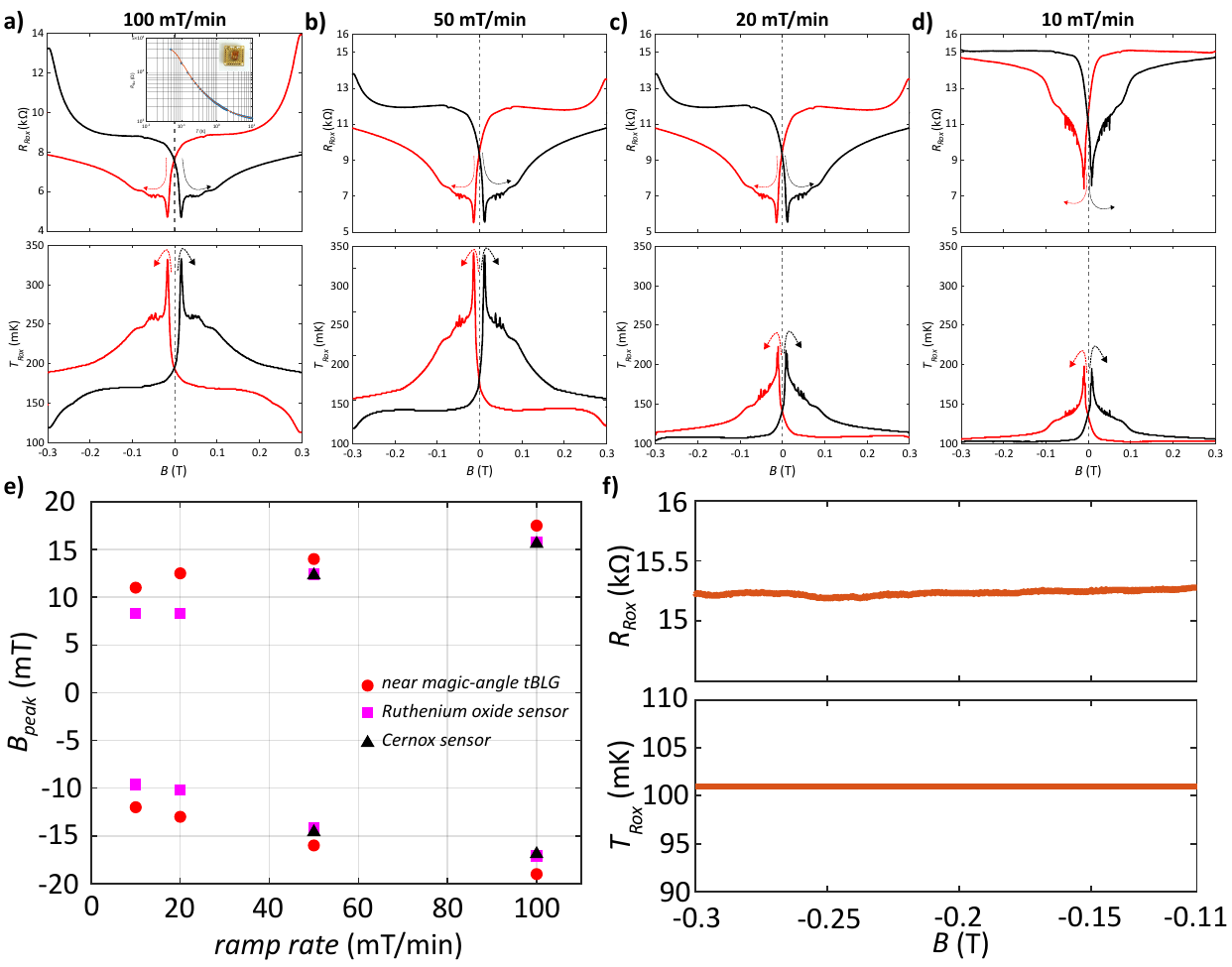}
	\caption[Hysteresis in  Ruthenium oxide temperature sensor with applied magnetic field $B$]{\textbf{Hysteresis in  Ruthenium oxide temperature sensor with applied magnetic field $B$:} \textit{Top} panel- $R^{FW}_{Rox}$ and $R^{BW}_{Rox}$ and \textit{Bottom} panel- equivalent temperature ($T^{FW}$ and $T^{BW}$) from the calibration curve shown in inset of \textbf{a)}. The ramp-rate was $100$ mT/min in \textbf{a)}, $50$ mT/min in \textbf{b)}, $20$ mT/min in \textbf{c)} and $10$ mT/min in \textbf{d)}. 
    \textbf{e)} $B_{peak}$ ($+$ve and $-$ve) for near magic-angle TBG, Ruthenium oxide sensor, and Cernox sensor for different ramp-rates of magnetic field. \textbf{f)} Ruthenium oxide sensor resistance/temperature in the steady state with a magnetic field variation with a very slow ramp-rate.} 
	\label{fig:4}
\end{figure*}

\begin{figure*}[htbp!]
	\centering
	\includegraphics[width=1.0\textwidth]{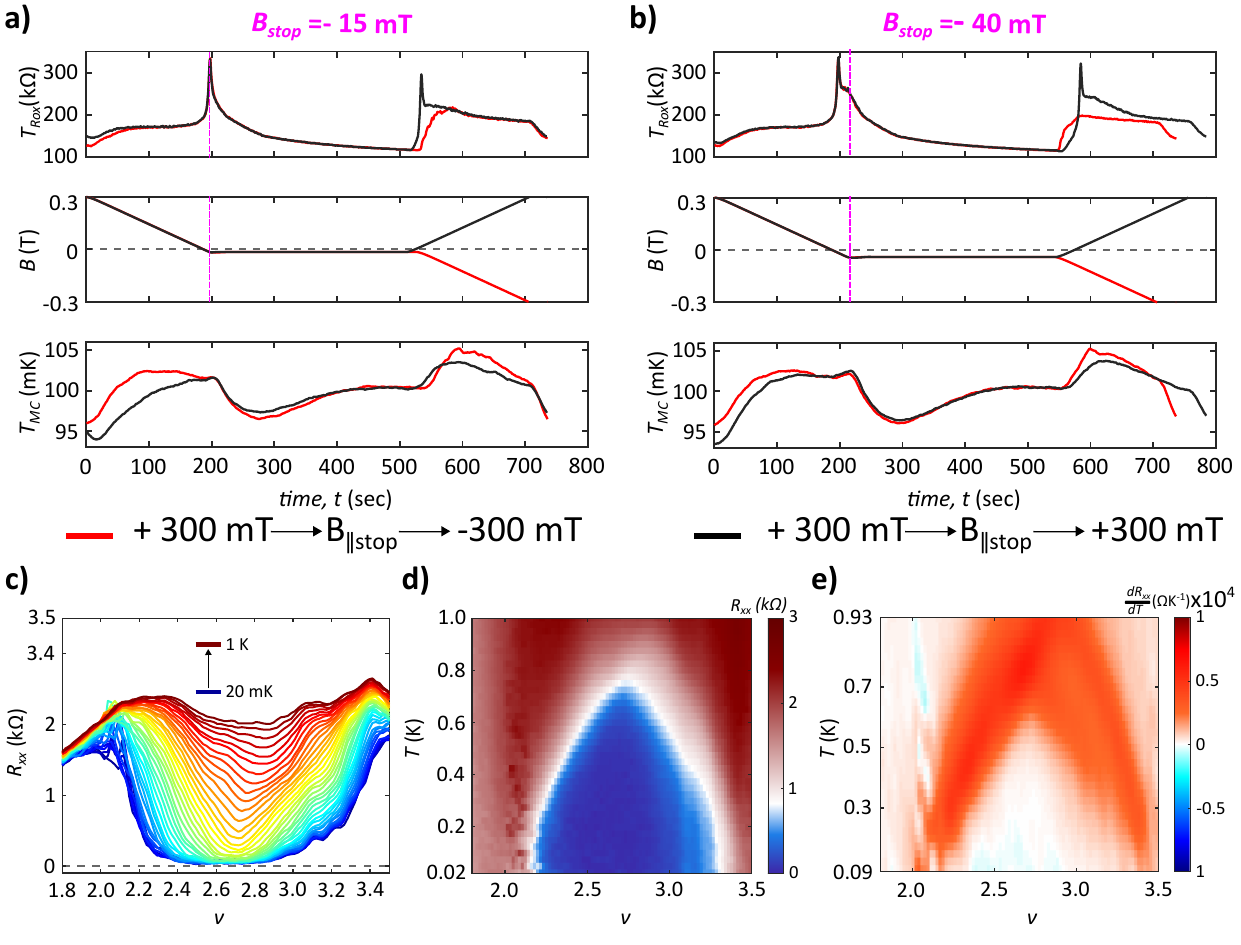}
	\caption[Transient experiment for hysteresis in  Ruthenium oxide sensor and effect of temperature on $R_{xx}$ in TBG]{\textbf{Transient experiment for hysteresis in  Ruthenium oxide sensor and effect of temperature on $R_{xx}$ in TBG:} Time-series plots of $T_{Rox}$, $B$ and $T_{MC}$ for \textbf{a)} $B_{stop}=-15$ mT and \textbf{b)}  $B_{stop}=-40$ mT with ramp-rate: 100 mT/min and excitation current, $I_s=10$ nA. \textbf{c)} Line plots of $R_{xx}$ vs. filling factors, $\nu$, with increasing $T$. \textbf{d)} $R_{xx}$($\nu,T$) $2-$d colormap, and \textbf{e)} ($\nu,T$) $2-$d colormap of temperature derivative of resistance ($\frac{d R_{xx}}{dT}$) at zero $D$.}
	\label{fig:5}
\end{figure*}

To check that, we loaded an independently calibrated Ruthenium oxide (RuO$_2$; Model no: RX-103A-BR) temperature sensor mounted on a chip carrier in the dilution fridge at the sample stage, which can measure temperature as low as $\sim50$ mK, closer to our operational base temperature of $70$ mK.
We subject the 
sensor to the exact condition of $B$ sweep in the forward and backward directions. We, surprisingly, get a BMR-like hysteresis for the $R_{Rox}$ as shown in Figure. \ref{fig:4}a (top panel) for a ramp-rate of $100$ mT/min. If converted to the temperature following the available $R$ vs. $T$ calibration curve of the sensor, we also find hysteresis in the $T_{Rox}$, in Figure. \ref{fig:4}a (bottom panel), similar to the `upwards' BMR presented for TBG in Figure. \ref{fig:1}d,f. The hysteresis in $R_{Rox}$ (or $T_{Rox}$) is observed for a magnetic field ramp-rate of as slow as $10$ mT/min (see Figure. \ref{fig:4}b-d). Following the observation of the effect of ramp-rate on the observed hysteresis in the $R_{xx}$ of TBG, we can compare the values of magnetic fields where the measured resistance of the TBG and sample stage temperature (from the sensor) show a peak with the 
magnetic field sweeps, \textit{i.e.}, $B_{peak}$. Figure. \ref{fig:4}e shows how the $|B_{peak}|$ values increase with increasing ramp-rate for both the TBG and the Ruthenium oxide sensor.
Surprisingly, $|B_{peak}|$ values are similar for the TBG and sensor. We have checked the same with an independently calibrated `Cernox' sensor, and the $B_{peak}$ for it at two different ramp rates are the same as that of the Ruthenium oxide sensor (\textcolor{magenta}{SI- 18}). This can re-enforce that the observed hysteresis in $R_{xx}$ might appear due to the mismatch between the temperature profile induced in the forward and backward magnetic field sweep at the sample stage, raising a question on the origin of the hysteresis observed around the SC phase in TBG could be an extrinsic effect rather than an intrinsic one. Similarly, if we perform the time-series experiment for measured $T_{Rox}$ (see Figure. \ref{fig:5}a,b), we observe the transient change in the measured temperature matching qualitatively with the transient behavior of $R_{xx}$,
giving us a glimpse of the change in profile of the local temperature at the sample stage.
Such rise and fall of the local sample stage temperature with the magnetic field sweep could arise with a possibility that the `sample puck' (see \textcolor{magenta}{SI-1}) is intrinsically susceptible to hysteresis loss or magnetocaloric effect.

Figure. \ref{fig:5}c shows $R_{xx}$ vs. $\nu$ around the SC region with increasing $T$ at $B=0$. From the $\nu-T$ $2-$d colormap of $\frac{dR_{xx}}{dT}$ (Figure. \ref{fig:5}e), we can see that at low temperature, $\frac{dR_{xx}}{dT}$
is high around $\nu\sim2$ and $\nu\sim3$, compared to $\nu$ closer to $\nu_{op}$ of the SC phase. As a result, $R_{xx}$ around $\nu\sim2$ and $\nu\sim3$ is highly sensitive to the local temperature change of the sample stage compared to the SC region, and $R_{xx}$ follows suit to the change in the temperature profile of the sample stage induced by magnetic sweep resulting in a hysteresis of the magnetoresistance. Quantitatively, for $R_{xx}$ hysteresis observed at $\nu= 2.1$ , $\Delta R^{max}_{xx}\approx0.55$ k$\Omega$ at $100$ mK (Figure. \ref{fig:1}d) and $\frac{dR_{xx}}{dT}\approx0.35\times10^4$ $\Omega$K$^{-1}$  at $100$ mK. So the change in $\Delta T$ due to continuous cycling of the magnet is $\sim160$ mK, comparable to the $\Delta T \approx200$ mK of the sample stage as measured from the Ruthenium oxide in Figure. \ref{fig:4}a for $\frac{dB}{dt}=100$ mT$/$min. Now for $T\geq T_c$, $\frac{dR_{xx}}{dT}$  increases vastly for $\nu_{op}-\delta < \nu_{op}$ and $ \nu_{op}+\delta>\nu_{op}$ making the region outside the superconductivity dome sensitive to the $\Delta T$ of the sample stage. This could be why we get finite $|\Delta R_{xx}|$ in Figure. \ref{fig:2}c for $T>400$ mK. An SC phase is increasingly more sensitive to a perpendicular magnetic field (\textcolor{magenta}{SI-Fig.  26}) with a rapid change in the magnetoresistance even with a small field overwhelming the effect of change of sample temperature compared to an applied $B_{||}$\cite{park2021tunable,su2023superconductivity,dutta2025electric}. The effect of local temperature change and the change in resistance in TBG is much more evident with the in-plane field, as seen in our experiment. Similarly, the hysteresis observed near zero filling ($\nu \sim 0$), with an opposite sign as shown in Figure. \ref{fig:2}a, is attributed to the negative slope of \(\frac{dR_{xx}}{dT}\) around the Dirac point, which is characteristic of its insulating behavior. In contrast, near full filling ($\nu \sim \pm 4$) in Figure. \ref{fig:2}a, the hysteresis remains unstable due to the high resistance values in that region. 



In summary, we observe a magnetoresistance hysteresis in the near magic-angle TBG around the SC phase under the application of an in-plane field. 
The shift in the observed $R_{xx}$ with varied ramp-rate and the anomaly between the transient nature of $R_{xx}$ in TBG under $B_{||}$ between two different sweep directions presents more questions than answers about the nature and origin of BMR in these systems. A close correspondence with the observed behavior of the measured local temperature of the sample stage under the applied field opens up a possibility to explore the effect of induced temperature through the application of a magnetic field, a possible extrinsic reason for observing such hysteresis in these systems, which require more careful experimental study to completely understand these findings and disentangle between the intrinsic and extrinsic sources. 

\end{spacing}

    


\section*{Associated Content}

\textcolor{magenta}{Supplementary Information} is available for this paper.

\section*{Data availability}
\begin{spacing}{2}
All the relevant non-analytical line-plot data generated or measured during this study are included in this published article (and its \textcolor{magenta}{Supplementary Information} files). Additional information related to this work is available from the corresponding author upon reasonable request.
\end{spacing}

\section*{Code availability}
\begin{spacing}{2}

The code that supports the findings of this study is available from the corresponding author upon reasonable request.
\end{spacing}

\textbf{Author contributions}\\
R.D. fabricated the device. R.D. and A.G. contributed to data acquisition and analysis. A.D. contributed to conceiving the idea and designing the experiment, as well as data interpretation and analysis. K.W. and T.T. synthesized the hBN single crystals. All the authors contributed to writing the manuscript.

\textbf{Notes}\\
The authors declare no competing financial interest.\\

\section*{Acknowledgements}
\begin{spacing}{2}

The authors thank Sharath Kumar and Deepshikha Jaiswal Nagar for sharing the calibrated temperature sensor. A.D. thanks the Department of Science and Technology (DST) and Science and Engineering Research Board (SERB), India, for financial support (SP/SERB-22-0387). A.D. also thanks CEFIPRA project SP/IFCP-22-0005. A.D. also acknowledges the funding support from the Department of Science and Technology (DST/NM/TUE/QM-5/2019), Government of India, under the Nanomission. Growing the hBN crystals received support from the Japan Society for the Promotion of Science (KAKENHI grant nos. 19H05790, 20H00354, and 21H05233) to K.W. and T.T.

\end{spacing}

\clearpage

\section{References}

\providecommand{\latin}[1]{#1}
\makeatletter
\providecommand{\doi}
{\begingroup\let\do\@makeother\dospecials
	\catcode`\{=1 \catcode`\}=2 \doi@aux}
\providecommand{\doi@aux}[1]{\endgroup\texttt{#1}}
\makeatother
\providecommand*\mcitethebibliography{\thebibliography}
\csname @ifundefined\endcsname{endmcitethebibliography}
{\let\endmcitethebibliography\endthebibliography}{}


 \newpage
\thispagestyle{empty}
\mbox{}


\section*{Supplementary material}

# Supplementary Information:

# Resistance hysteresis in twisted bilayer graphene: Intrinsic versus extrinsic effects

Ranit Dutta[1]*, Ayan Ghosh[1]†, Kenji Watanabe[2], Takashi Taniguchi[3], and Anindya Das[1]‡

[1]Department of Physics, Indian Institute of Science, Bangalore, 560012, India

[2]Research Center for Functional Materials, National Institute for Materials Science, 1-1 Namiki, Tsukuba 305-0044, Japan.

[3]International Center for Material Nanoarchitectonics, National Institute for Materials Science, 1-1 Namiki, Tsukuba 305-0044, Japan

# Contents

*equally contributed

†equally contributed

‡anindya@iisc.ac.in



## SI- 1 : Loading device for experiment with an in-plane magnetic field ($B_{||}$)

We have performed measurements with $B_{||}$ in a cryo-free dilution refrigerator fitted with a superconducting magnet. Though there are (dilution) fridges available commercially that come with a magnet that can be applied in all three cartesian directions, $X, Y$, and $Z$, we were limited to a setup that can generate field lines only in one direction (the proverbial $Z-$ axis or out of plane). To overcome this limitation, our device was loaded on a custom-made chip holder that essentially rotates the plane of the device by $90°$ while loading, such that now the plane of the device becomes parallel with the $Z-$axis as shown in SI-Fig. 1a,b. The whole arrangement can now be loaded on the cold finger (SI-Fig. 1c,d) sitting inside a 'sample puck', which can then be bottom loaded in the dilution fridge. In this arrangement, the magnetic field lines run parallel to the sample puck's and cold finger's central axis.

## SI- 2 : Butterfly Magnetoresistance (BMR) - A discussion

BMR are a cross magnetoresistance (MR) hysteresis curves resembling the shape of wings of a butterfly and observed in a diverse range of magnetic systems including but not limited to van der Waals materials, strongly correlated systems, and traditional magnets [1–4] along with other ubiquitously observed traditional MR hysteresis curves, for example, in the shape of a parallelogram. An MR curve reflects the evolution path of the magnetic structure of a system when subjected to an external varying magnetic field. An MR curve becomes a BMR when the MR passes through two higher (lower) resistance states in the sweeping-forward and sweeping-backward processes of the applied magnetic field. The MR curve displays two crossed peaks (valleys) and results in an upward (downward) BMR, as shown in SI-Fig. 2a,b. As seen from the schematics, for large enough +ve and −ve B (> coercive field, $|B_{co}|$), magnetoresistance is almost similar, indicating an ordered phase of a magnetic system. It is evident from SI-Fig. 2a,b, the peaks (valleys) appear during the switching of the magnetization direction with a sweep of the magnetic field, moving the ordering through two volatile states (one in forward sweep and another in backward sweep), resulting in hysteresis between MR curves. As we are talking about layered 2D materials, it is imperative to consider the anisotropy of the system, *i.e.*, the direction of the easy/hard axis. The arrangement of magnetic ordering favoured by a particular anisotropy can lead to resistance values lower or higher than the volatile states for $|B| > |B_{co}|$. For a given anisotropy of a system, it is possible to change/switch the magnetic anisotropy by subjecting the system to various external experimental perturbations like applied current, temperature, and electric field. The implication of this is, that a magnetic system showing BMR hysteresis can be switched from a magnetic ordering with 'upwards' BMR (SI-Fig. 2a) to a magnetic ordering 'downwards' BMR (SI-Fig. 2b) or vice versa under appropriate tuning of experimental conditions/parameters.

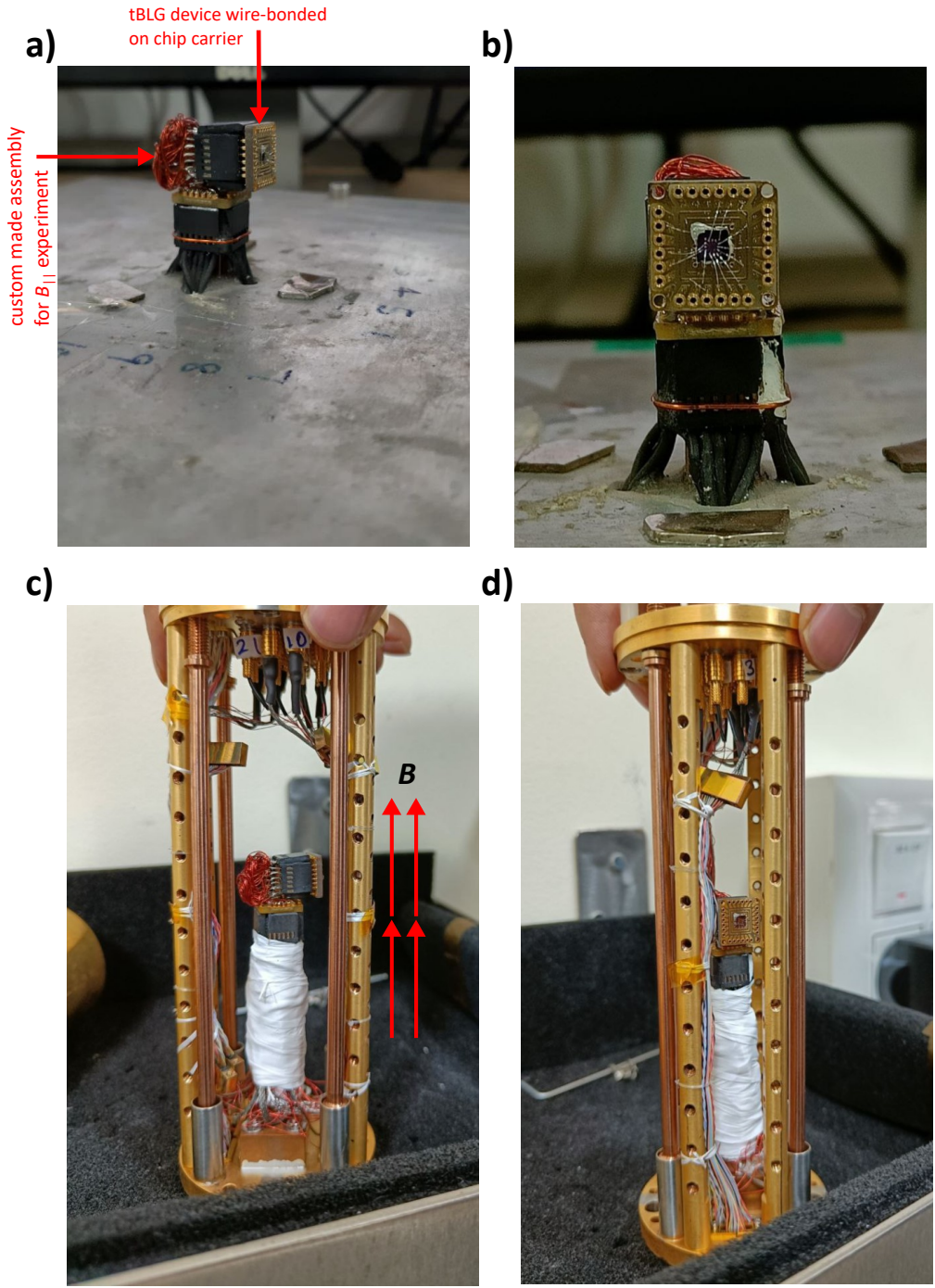

**SI-Fig. 1: Loading device for $B_{||}$ experiment in dilution fridge: a)** Side and **b)** Front view of the wire bonded device on a chip carrier loaded on a homemade chip holder. In this arrangement, the device plane is parallel to the '$Z$'− axis. **c)** Side and **d)** The front view of the arrangement is attached to the cold finger sitting on a 'sample puck' to load in the dilution fridge through the bottom loader. The magnetic field lines generated by the superconducting magnet run parallel to the central axis of the sample puck/cold finger (schematically shown by red arrows), making them parallel to the device plane.

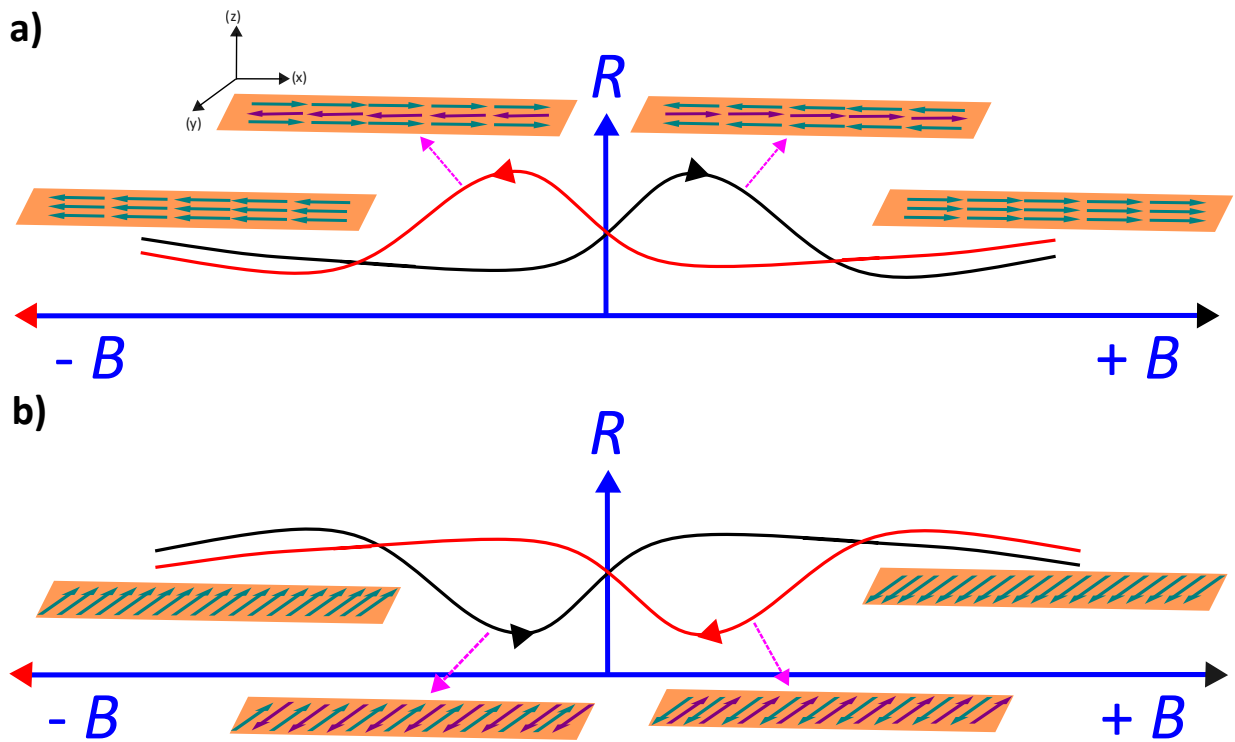

**SI-Fig. 2: BMR:** Schematic MR curves resembling the shape of the wings of a butterfly and 'toy model' for magnetic ordering shown with magnetic moments aligned along a preferential direction assuming the presence of magnetic anisotropy for a **a)** upwards and **b)** downwards BMR hysteresis curves.

## SI- 3 : Effect of temperature of the Mixing Chamber (MC) plate with $B_{||}$ sweep

The measured TBG sample is connected to a cold finger made out of copper sitting inside a sample puck as we have shown in SI-Fig. 1 (section SI- 1). The sample is cooled to the measurement (base) temperature through the wirings around the cold finger to thermally equilibrate to the mixing chamber (MC) plate temperature of the cryo-free dilution refrigerator connecting the leads (legs) of the chip carrier to the MC plate of the dilution fridge. The 'bath' temperature or the MC plate temperature could be read out through a Ruthenium oxide (Rox$^{TM}$) temperature sensor. It has been seen that, there is an increase in the bath (MC plate) temperature, $T_{MC}$, whenever the magnetic field is swept/ramped in either direction of the $B$ fields. For example, say we switch on the magnet and ramp up to $+300$ mT with a ramp rate of $100$ mT/min, $T_{MC}$ rises to $\sim 72$ mK from the base operation temperature of $\sim 20-25$ mK. Throughout this report, all the data presented here for $B_{||}$ sweep hysteresis has been obtained for the ramp rate of $100$ mT/min unless otherwise mentioned. We have tried to see if such a change of $T_{MC}$ with a continuous $B_{||}$ sweep has any effect on the observed nature of BMR response and $\Delta R_{xx}$ magnitude at different $\nu$. In the similar spirit to $R_{xx}^{FW}$ and $R_{xx}^{BW}$ (as defined in main text), $T_{MC}^{FW}$ and $T_{MC}^{BW}$ are plotted in SI-Fig. 3 for different values of $\nu$. $T_{MC}^{FW}$ and $T_{MC}^{BW}$ are the profile of the MC plate temperature, $T_{MC}$, for forward and backward $B_{||}$ sweep directions respectively. The $B_{||}$ hysteresis loop path is traced as $+B_{||} \to -B_{||}$ [$i) \to ii)$], see SI-Fig. 3a (bottom panel)) and then $-B_{||} \to +B_{||}$ [$ii) \to iii)$] starting from $B_{||} = +300$ mT. As shown in the bottom panels of SI-Fig. 3a-c, the $T_{MC}$ profile for the 'forward' and 'backward' sweep are always similar for any choice of $\nu$ irrespective of the nature of BMR (hysteresis or no hysteresis). Also, the temperature increase profile is monotonic, which does not correlate to the non-monotonic increase or decrease of $R_{xx}$ vs. $B_{||}$ curves. We would also like to point out that the change in $T_{MC}$ is relatively small, $\Delta T \sim 10 - 15$ mK.

## SI- 4 : Evolution of hysteresis phase with temperature

From the observations presented in Figure 2a of the main text, we see that the hysteresis phase resides at the boundary of the superconducting phase observed in this near magic-angle TBG. Also, when the SC phase can be tuned to a normal phase by changing the temperature above a critical temperature, $T_c$, a weak hysteresis phase emerges from the region deep inside the SC phase (Figure 2c of the main text). The $T_c$ for the SC phase for the near magic-angle TBG is $\sim 800$ mK at $\nu_{op} \sim 2.7$ at zero $D$ $^?$.

As shown in SI-Fig. 4, we plot $\Delta R_{xx}$ with $B_{||}$ for zero $D$ around the SC region ($\nu \sim 1.5 - 3.5$) with increasing sample/bath temperature from $70$ mK to $1$ K. The two hysteresis 'islands' start weakening and move closer towards each other with increasing $T$ and above $T \approx 600$ mK merge in a single hysteresis phase in the filling range $\nu \sim 2.2 - 3$. These observations again highlight the emergence of the hysteresis phase with a gradual weakening of the SC phase achieved through the increase of the sample (TBG) temperature above $T_c$.

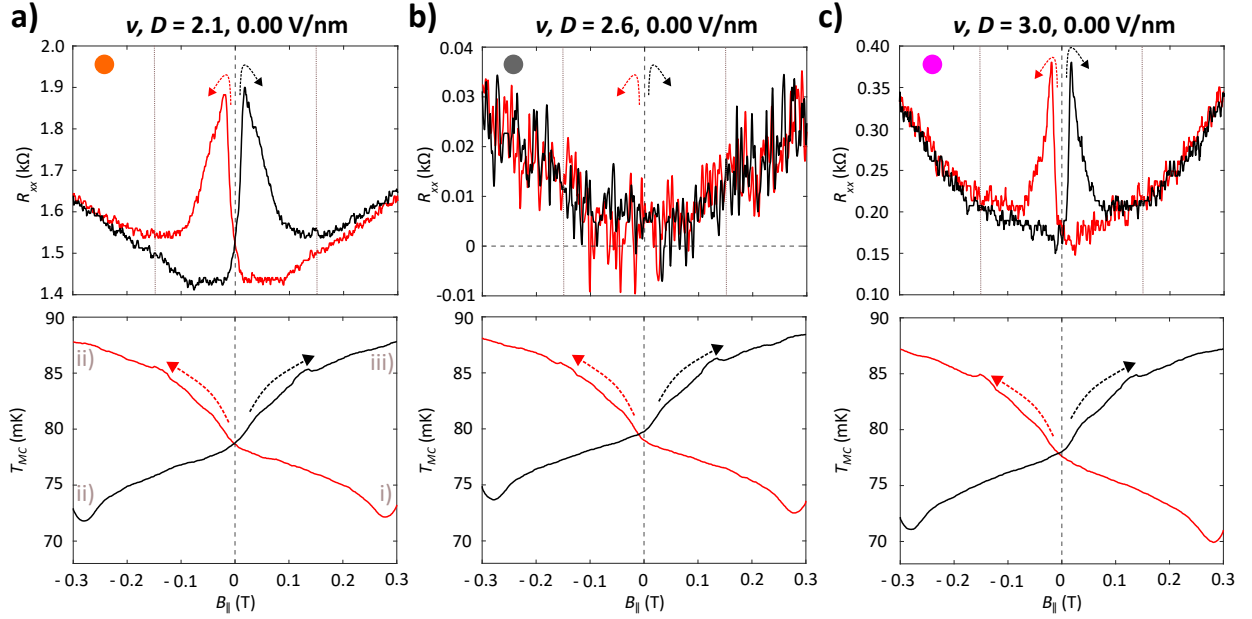

**SI-Fig. 3:** $R_{xx}$ **and measured** $T_{MC}$ **profile with forward and backward** $B_{\parallel}$ **sweep:** BMR hysteresis traces ($R_{xx}^{FW}$ and $R_{xx}^{BW}$) (top panel) and mixing chamber plate/bath temperature profile ($T_{MC}^{FW}$ and $T_{MC}^{BW}$) (bottom panel) within the $B_{\parallel}$ sweep range of $\pm 300$ mT for **a)** $\nu = 2.1$, **b)** $\nu = 2.6$, **c)** $\nu = 3.0$ (at zero $D$). The $B_{\parallel}$ sweep rate is: 100 mT/min. At the start of the hysteresis loop $T_{MC}^{FW}$ rises continuously from $\sim 72$ mK to $\sim 90$ mK as $B_{\parallel}$ is changed continuously from $+300$ mT to $-300$ mT. We add a delay of 2 minutes before we switch back to the other sweep direction. Due to this, $T_{MC}$ decreases to $\sim 75$ mK, closer to the initial temperature at the beginning of the hysteresis loop. The rise of $T_{MC}^{BW}$ is similar with the opposite direction of $B_{\parallel}$ sweep. We see that even if there is a temperature difference of $\Delta T_{MC} \sim 10$ mK between the two sweep directions for $|B_{\parallel}| \sim 150 - 300$ mT, $R_{xx}^{FW} \approx R_{xx}^{BW}$ for all values of $\nu$ for $|B_{\parallel}| > 150$ mT.

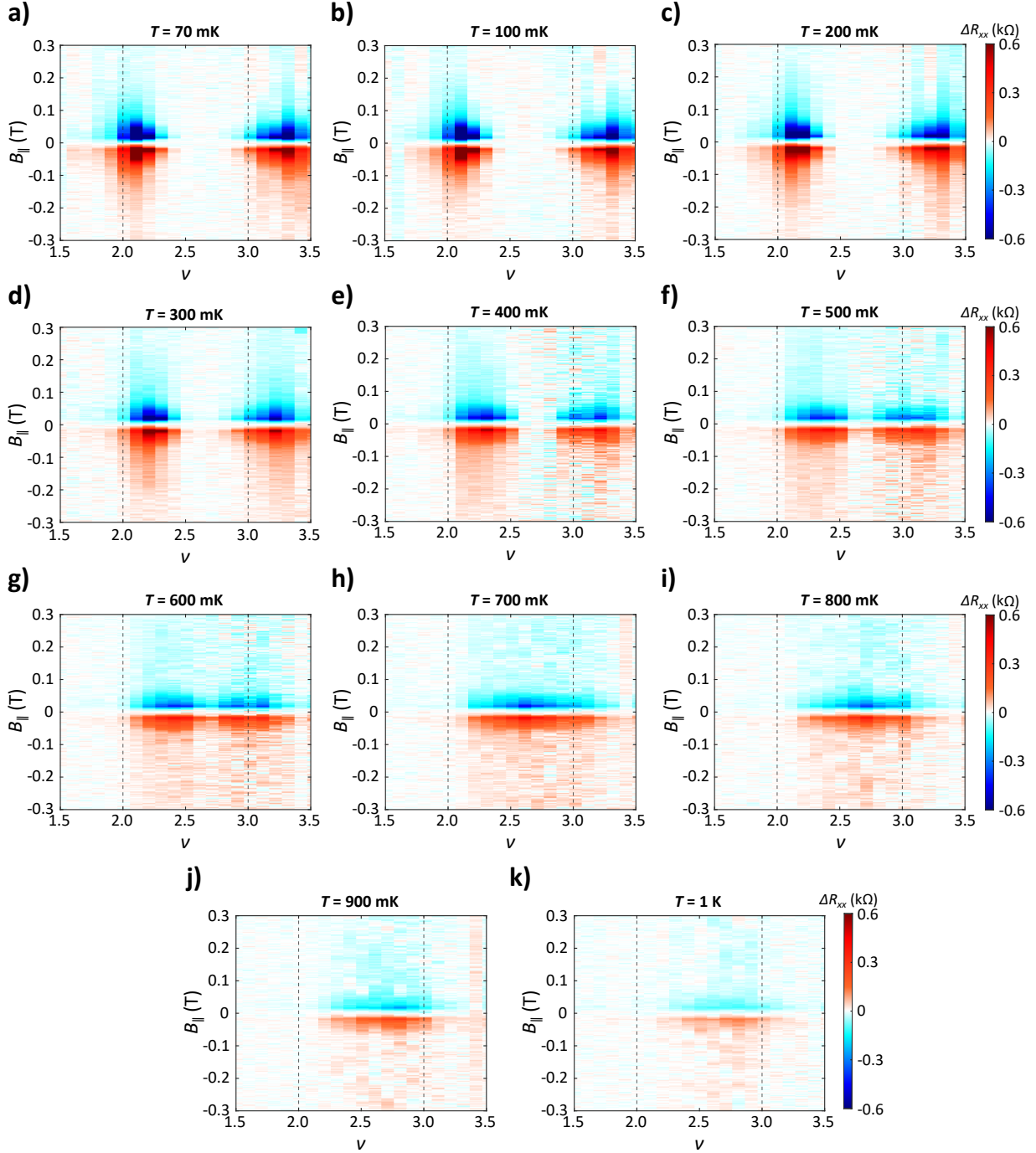

**SI-Fig. 4: Temperature evolution of hysteresis islands at zero $D$:** $\Delta R_{xx}(\nu, B_{||})$ 2−d colormap at $D =$ 0.00 V/nm at sample temperatures of **a)** 70 mK, **b)** 100 mK, **c)** 200 mK, **d)** 300 mK, **e)** 400 mK, **f)** 500 mK, **g)** 600 mK, **h)** 700 mK, **i)** 800 mK, **j)** 900 mK, and **k)** 1 K. The two hysteresis islands centred around $\nu \sim 2.2$ and $3.2$ at the lowest temperature gradually weaken and merge to form a singular phase above 600 mK.

## SI- 5 : Transient effect of the hysteresis response in TBG

As shown in Figure 3b of the main text, measured longitudinal magnetoresistance transiently relaxes to a steady state magnitude of $R_{xx}$ when we stop sweeping the magnetic field at a given $B_{||stop}$. As highlighted in the main text, the transient recovery behavior after the 'stop interval' are markedly different for two different sweeping protocols. Below we elaborate on this behavior for different $B_{||stop}$ values.

### I. Transient/ Time series experiment protocol in hysteresis

1. Choosing a point in the accessible filling factor, $\nu$, phase space where we see a fully developed BMR response, *i.e.*, magnetoresistance hysteresis, with a significant magnitude of $\Delta R_{xx}$.

2. We magnetize and de-magnetize the sample by performing a closed loop cycle of $+B_{||} \rightarrow -B_{||}$ ($+300$ mT to $-300$ mT) and $-B_{||} \rightarrow +B_{||}$ ($-300$ mT to $+300$ mT).

3. We start ramping the magnet from $B_{||} = +300$ mT towards $-300$ mT, but instead of completing the forward sweep, we stop the magnet ramping at different magnetic field values designated by $B_{||stop}$.

4. We sit at $B_{||stop}$ for 300 seconds ('stop interval') and keep measuring continuously $R_{xx}$ with time, $t$, at that $\nu$ with the applied in-plane magnetic field of $B_{||stop}$ during that interval.

5. After the interval of 300 seconds the magnet is again started ramping; this can be performed in two ways: (i) we ramp backward, *i.e.*, $B_{||stop} \rightarrow +300$ mT, (ii) we can complete the forward sweep from $B_{||stop} \rightarrow -300$ mT.

Below we discuss the results of these transient experiments.

### II. Transient behavior of $R_{xx}$ in $+300$ mT $\rightarrow B_{||\mathbf{stop}} \rightarrow +300$ mT

SI-Fig. 5a-f show the results for the time series measurement for different $B_{||stop}$ values at $\nu = 2.2$. We have plotted the measured $R_{xx}$ (top panel), $B_{||}$ (middle panel), and temperature of dilution MC plate, $T_{MC}$ (bottom panel) with time, $t$, adding the 300 second stop interval for each $B_{||stop}$. The inset in the $R_{xx}$ vs. $t$ plots show the $R_{xx}$ vs. $B_{||}$ trace along the sweep direction of $+300$ mT $\rightarrow B_{||stop} \rightarrow +300$ mT. SI-Fig. 5a shows $R_{xx}$, $B_{||}$ and $T_{MC}$ with time, $t$, for $B_{||stop} = 0$. We see that even though in the continuous sweep $R_{xx}$ follows the path of $R_{xx}^{FW}(B_{||})$ (at a given $\nu$) continuously, as soon as we stop ramping and hold the magnet at $B_{||stop}$, $R_{xx}$ quickly starts decreasing and settles at a steady state value [5] in a very short time scale within the stop interval of 300 seconds. When we re-start ramping the magnet in the $B_{||stop} \rightarrow +300$ mT direction, we recover a peak in the $R_{xx}$. We observe that the magnitude of $R_{xx}$ peak depends on the $B_{||stop}$,

but the value of $B_{||}(+$ ve$)$ where it appears is curiously around $\sim 17-18$ mT, similar to the position of the peak in $R_{xx}^{BW}$ as shown in SI-Fig. 5 and 14a.

**III. Transient behavior of $R_{xx}$ in $+300$ mT $\to B_{||\text{stop}} \to -300$ mT**

We repeat the same thing with similar choices of $B_{||stop}$, but with ramping the magnet towards $-300$ mT after the stop interval, *i.e.*, $+300$ mT$\to B_{||stop} \to -300$ mT. The results are demonstrated in SI-Fig. 6a-f. We have plotted the measured $R_{xx}$ (top panel), $B_{||}$ (middle panel), and temperature of dilution MC plate, $T_{MC}$ (bottom panel) with time, $t$, adding the 300 second stop interval for each $B_{||stop}$. The inset in the $R_{xx}$ vs. $t$ plots show the $R_{xx}$ vs. $B_{||}$ trace along the sweep direction of $+300$ mT $\to B_{||stop} \to -300$ mT. SI-Fig. 6a shows $R_{xx}$ (top panel), $B_{||}$ (middle panel) and $T_{MC}$ (bottom panel) with time, $t$, for $B_{||stop}=0$. Just like we discussed in the previous section, We see that even though in the continuous sweep $R_{xx}$ follows the tath of $R_{xx}^{FW}(B_{||})$ (at a given $\nu$) continuously, as soon as we stop ramping and hold the magnet at $B_{||}stop$, $R_{xx}$ quickly starts decreasing and settles at a steady state value [5] in a very short time scale during the stop interval of 300 seconds, irrespective of the value of $B_{||stop}$. After the stop interval, as soon as we start ramping $B_{||}$ again towards $B_{||}=-300$mT, we see the appearance of a peak around $B_{||} \sim -17.2$ mT. Now, for $-40$ mT $< B_{||stop} < 0$, $R_{xx}$ trace develops a small local maxima with magnet sweep after the stop interval before recovering the typical values of $R_{xx}^{FW}$ (see SI-Fig. 6 and 14a) with $B_{||}$ sweep towards $-300$ mT.

SI-Fig.7 shows a comparison of the transient effect observed in $R_{xx}$ for the two sweep protocols/ schemes introduced in section SI- 5.

**SI- 6 : Current driven ($I_{DC}$) switching in reduced differential magnetoresistance (rDMR)**

Near the magic angle, TBG hosts a stable superconducting phase near zero $D$ around $\nu \sim 2-3$ (see main text). In SI-Fig. 8, we have plotted differential resistance, $dV_{xx}/dI$, as a function of D.C. bias current, $I_{DC}$, for different filling factors around the SC region. Near the optimal doping, $\nu_{op} \sim 2.7$, we get $dV_{xx}/dI \approx 0$ below the critical current, $I_c$, before rising sharply and transitioning from a SC state to a normal state. The differential resistance behavior is in striking contrast with $dV_{xx}/dI$ vs. $I_{DC}$ for other $\nu$ values where superconductivity is not present.

**Hysteresis in presence of applied $I_{DC}$ under $B_{||}$**

SI-Fig.s 9a,c,e show how differential resistance, $dV_{xx}/dI$, changes with increasing $B_{||}$ up to $+300$ mT for $|I_{DC}| \geq 0$ for three representative filling factors around the SC region. With increasing $B_{||}$, $dV_{xx}/dI$ is also seen to increase upto a certain value of $I_{DC}$ before the behavior is switched (defined as $I_s$), *i.e.*, $dV_{xx}/dI$ decreases with increasing $B_{||}$. We can perform the hysteresis measurement at a fixed value of

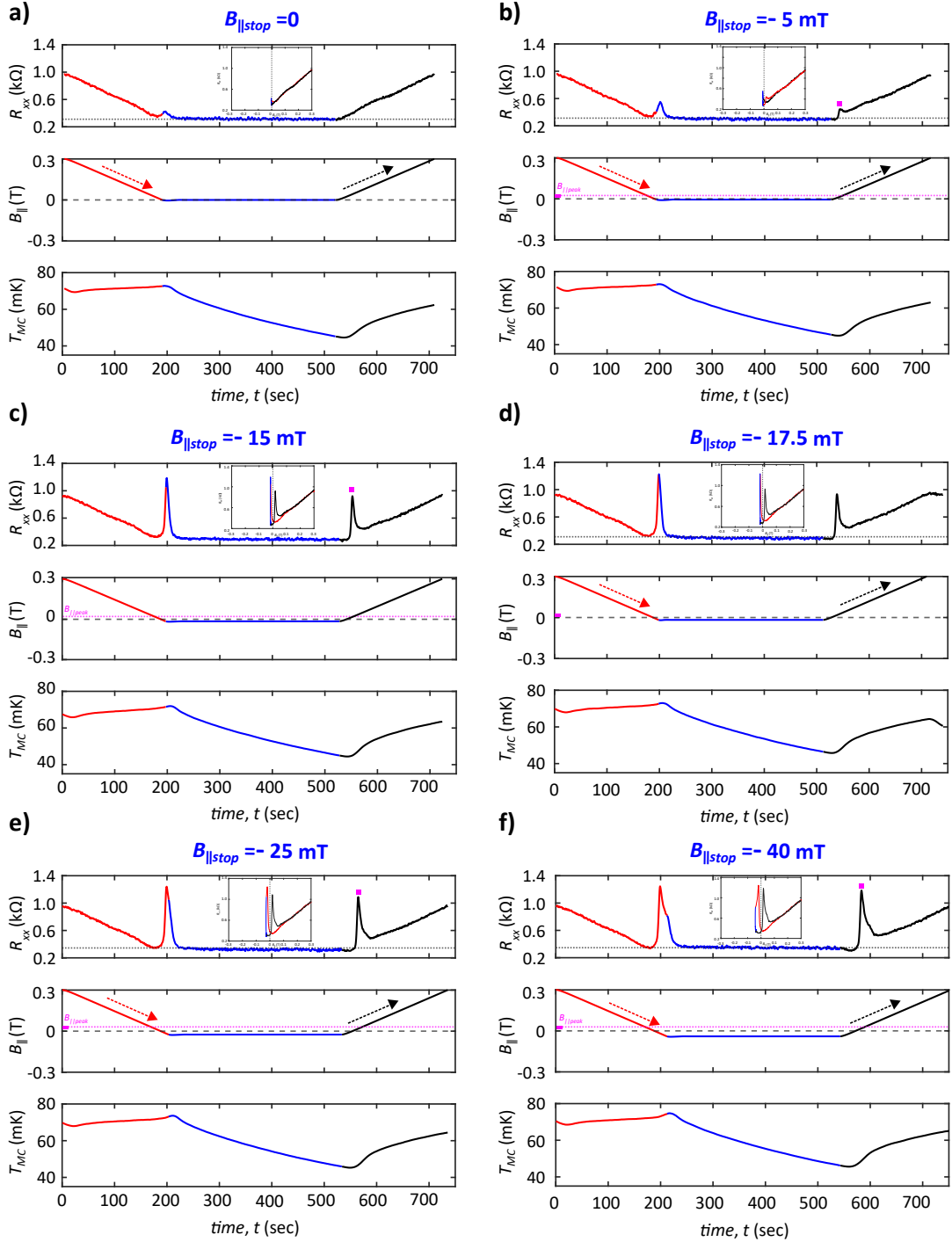

**SI-Fig. 5: Transient behavior of $R_{xx}$ in $+300$ mT $\to B_{\|stop} \to +300$ mT at $(\nu, D) = 2.2, 0.00$ V/nm:** $R_{xx}$ (top panel), $B_{\|}$ (middle panel), and $T_{MC}$ (bottom panel) with time, $t$, plotted in three panels for $B_{\|stop} =$ **a)** 0, **b)** $-5$ mT, **c)** $-15$ mT, **d)** $-17.5$ mT, **e)** $-25$ mT, and **f)** $-40$ mT. Inset in each of the top panels show the $R_{xx}$ vs. $B_{\|}$ trace for $+300$ mT $\to B_{\|stop} \to +300$ mT. $R_{xx}$ transiently relaxes to (blue trace) a steady state for all the $B_{\|stop}$ just at the beginning of stop interval of 300 seconds. Red arrow (↘) in the middle panels indicates to the magnetic sweep direction in $+300$ mT to $B_{\|stop}$ and the **Black** arrow (↗) indicate to the magnetic sweep direction in $B_{\|stop}$ to $+300$ mT. The dotted line in the middle panels corresponds to the value of $B_{\|}$ ($B_{\|peak}$) at which the peak in $R_{xx}$ re-appears after the stop interval. These $B_{\|}$ values are $+16.3$ mT for (**b**), $+17$ mT for (**c**), $+17.1$ mT for (**d**), $+18$ mT for (**e**), and $+18.3$ mT for (**f**).

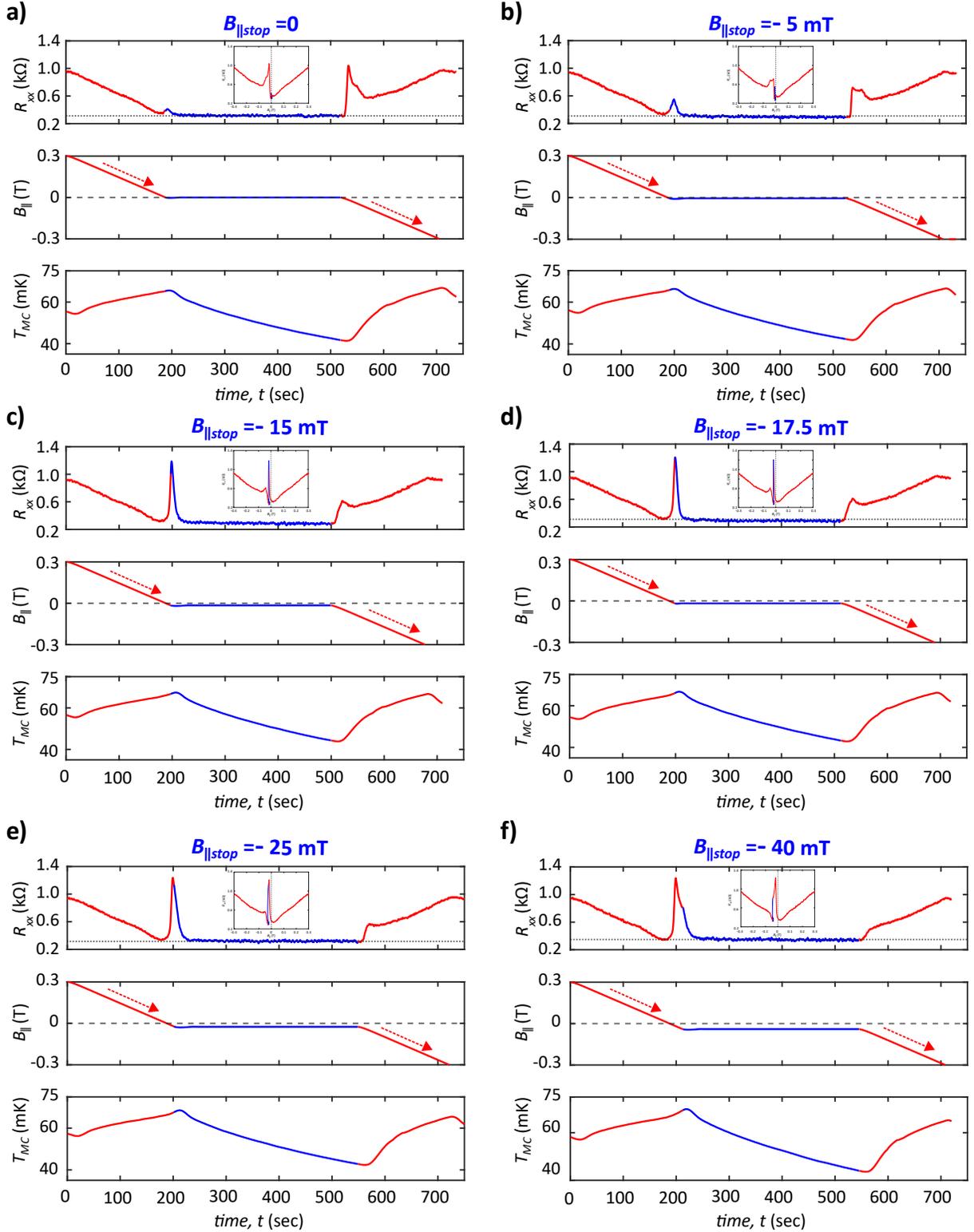

SI-Fig. 6: **Transient behavior of $R_{xx}$ in $+300$ mT $\to B_{\|\text{stop}} \to -300$ mT at $(\nu, D) = 2.2, 0.00$ V/nm:** $R_{xx}$, $B_{\|}$, and $T_{MC}$ with time, $t$, plotted in three panels for $B_{\|stop} =$ **a)** $0$, **b)** $-5$ mT, **c)** $-15$ mT, **d)** $-17.5$ mT, **e)** $-25$ mT, and **f)** $-40$ mT. Inset in each of the top panel show the $R_{xx}$ vs. $B_{\|}$ trace for $+300$ mT $\to B_{\|stop} \to -300$ mT. $R_{xx}$ transiently relaxes (blue trace) to a steady state for all the $B_{\|stop}$ just at the beginning of stop interval of 300 seconds. Red arrow (↘) in the middle panels indicate the magnetic sweep direction in $+300$ mT to $-300$ mT.

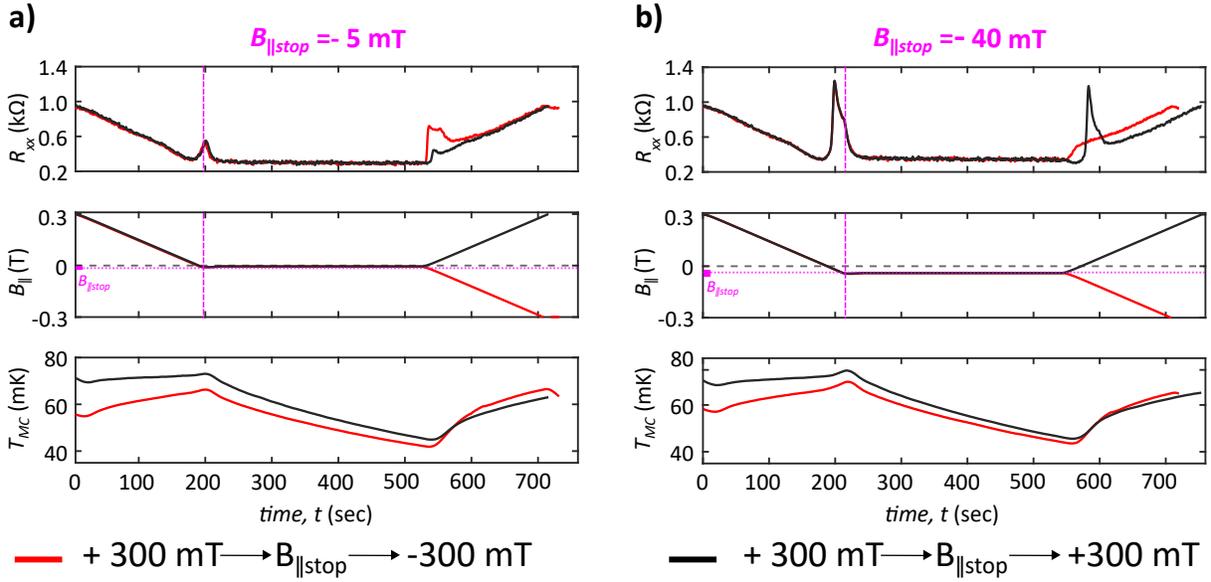

**SI-Fig. 7: Transient effect of $R_{xx}$ in hysteresis in TBG:** (From top to bottom) $R_{xx}$, $B_{||}$ and $T_{MC}$ transient sweep with time, $t$, plotted in three panels for $B_{||stop} =$ **a)** $-5$ mT and **b)** $-40$ mT at $\nu = 2.2$.

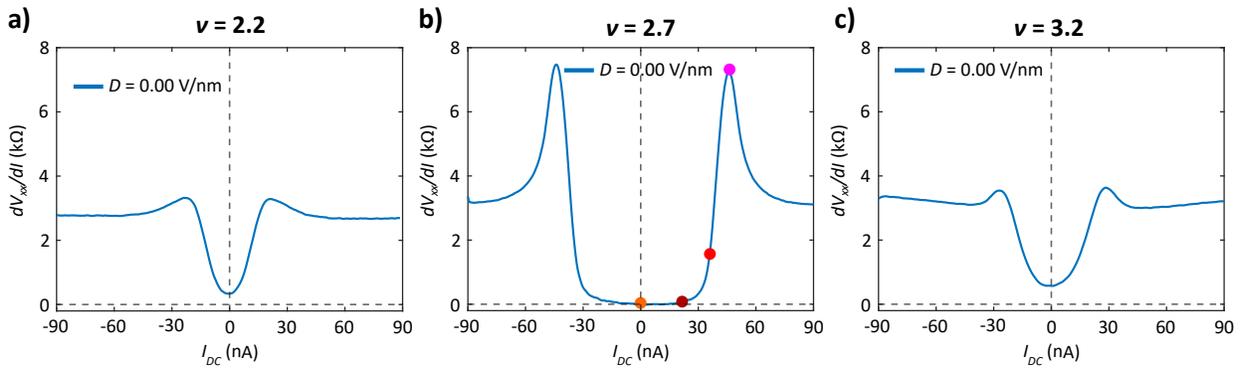

**SI-Fig. 8: $dV_{xx}/dI$ vs. $I_{DC}$:** Differential resistance is plotted as a function of applied $I_{DC}$ for $D = 0.00$ V/nm at filling factors of $\nu =$ **a)** 2.2, **b)** 2.7, and **c)** 3.2. Solid circles on the $dV_{xx}/dI$ curve of **(b)** are at different $I_{DC}$ values: 0 (●), 21.60 nA (●), 36 nA (●), and 45 nA (●).

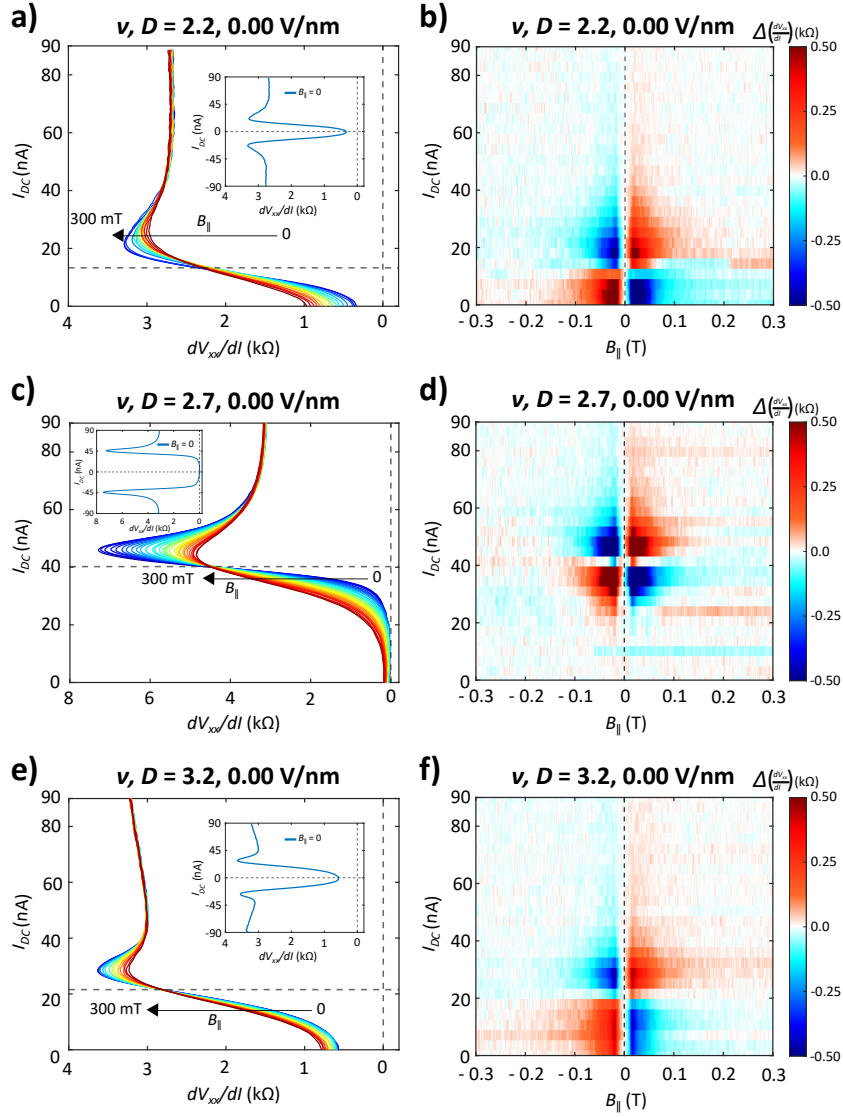

**SI-Fig. 9: Sign reversal of rDMR at zero $D$ with $I_{DC}$:** $\frac{dV_{xx}}{dI}$ plotted as a function of $I_{DC}$, with increasing in-plane magnetic field from 0 to 300 mT for $D = 0.00$ V/nm at $\nu$ values of **a)** 2.2, **c)** 2.7, and **e)** 3.2. $dV_{xx}/dI$ traces have been shown only for positive values of $I_{DC}$. The behavior are symmetric across zero current for $I_{DC} < 0$. Inset shows the $dV_{xx}/dI$ vs. $I_{DC}$ for the corresponding filling factors at $B_{||} = 0$. We observe that, initially $dV_{xx}/dI$ increases with increasing $B_{||}$ but the behavior reverses ($dV_{xx}/dI$ decreases with increasing $B_{||}$) at a sufficiently higher value of $I_{DC}$ defined as $I_s$. The $I_s$ values for the three filling factors are $\sim 14$ nA ($\nu = 2.2$), $\sim 40.5$ nA ($\nu = 2.7$), and $\sim 22.5$ nA ($\nu = 3.2$). The reversal in the nature $dV_{xx}/dI$ with applied $B_{||}$ below and above $I_s$ is also reflected in the reduced differential magnetoresistance (rDMR). Below the $I_s$, BMR response resembles the nature shown in SI-Fig. 14a, and above $I_s$ the BMR response is flipped like in SI-Fig. 14c. As a result, we also see a reversal of sign in rDMR for a given $B_{||}$ with increasing $I_{DC}$ as seen in the 2−d colormap of $\Delta\frac{dV_{xx}}{dI}(B_{||}, I_{DC})$ for $\nu =$ **b)** 2.2, **d)** 2.7, and **f)** 3.2. The values of $I_{DC}$ where rDMR shows the sign reversal in the colormaps match very closely to the $I_s$ values for each $\nu$ found from the line plots of $dV_{xx}/dI$ vs. $I_{DC}$ with increasing $B_{||}$.

applied $I_{DC}$ by measuring $dV_{xx}^{FW}/dI$ and $dV_{xx}^{BW}/dI$ with forward and backward $B_{||}$ sweep and calculate the reduced differential MR or rDMR defined as:

$$\Delta\frac{dV_{xx}}{dI} = dV_{xx}^{FW}/dI(+B_{||} \to -B_{||}) - dV_{xx}^{BW}/dI(-B_{||} \to +B_{||}) \quad (1)$$

We find that if the applied $I_{DC}$ value is smaller or greater than $I_s$ for a choice of $\nu$, we get a reversal of sign in rDMR. To put it simply, for a $B_{||}$ value (+300 mT $< B_{||} <$ −300 mT) we can cross over from positive (negative) to negative (positive) rDMR as we change applied D.C. bias current from $I_{DC} < I_s$ to $I_{DC} > I_s$ as demonstrated in $\Delta\frac{dV_{xx}}{dI}(B_{||}, I_{DC})$ 2−d colormaps in SI-Fig.s 9b and f before the collapse of rDMR at a higher bias current ($I_{DC} >$ 50 nA). At the optimal doping of $\nu_{op} = 2.7$, we don't see any sign of hysteresis ($\Delta\frac{dV_{xx}}{dI} \approx 0$; SI-Fig. 9d) below a bias current of $\sim$ 30 nA, in line with the observations already described in the main text. Once SC region starts transitioning to the normal state with increasing D.C. bias current, i.e., $\frac{dV_{xx}}{dI} > 0$, we see a hysteresis phase emerging at the optimal doping, the nature of which is switched for $I_{DC} > I_s$, where $I_s \approx 38$ nA for $\nu = 2.7$.

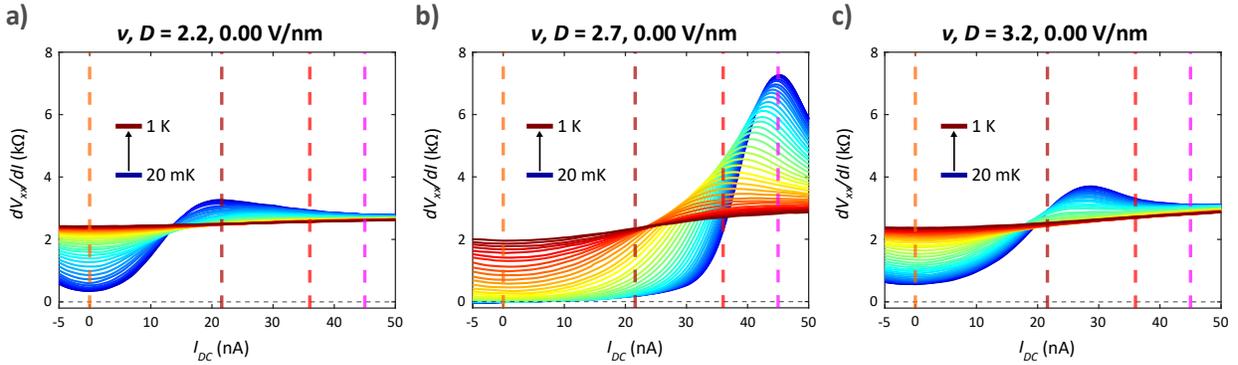

**SI-Fig. 10:** $\frac{dV_{xx}}{dI}$ **vs.** $I_{DC}$ **with** $T$**:** $\frac{dV_{xx}}{dI}$ plotted as a function of applied $I_{DC}$ with increasing $T$ from 20 mK to 1 K for $D = 0.00$ V/nm at $\nu$ values of **a)** 2.2, **b)** 2.7, and **c)** 3.2. Dashed vertical lines are at $I_{DC}$ values of 0, 21.60 nA, 36 nA, and 45 nA. $dV_{xx}/dI$ increases with increasing $T$ but the behavior reverses ($dV_{xx}/dI$ decreases with increasing $T$) at a sufficiently higher value of $I_{DC}$ defined as $I_s$. The $I_s$ values for the three filling factors are $\sim$ 14 nA ($\nu = 2.2$), $\sim$ 41 nA ($\nu = 2.7$), and $\sim$ 23 nA ($\nu = 3.2$).

Now, how do we understand such switching of the nature and the sign in rDMR (or rMR) with increasing $I_{DC}$. As discussed in the main text (Figure 5) the continuous change in the magnetic field might lead to an induced non-monotonous change and hysteresis in the local temperature profile of the sample stage (and sample) in the forward and backward magnetic sweep, and depending on the nature of $R_{xx}$ vs. $T$, at a given $\nu$, the sign of rMR is determined. When $\frac{dR_{xx}}{dT}|_\nu > 0$, $R_{xx}(\nu)$ increases with increasing $T$, and we get positive (negative) $\Delta R_{xx}$ for negative (positive) $B_{||}$ (see Figure 3 in the main text). On the contrary, if $\frac{dR_{xx}}{dT}|_\nu < 0$, $R_{xx}(\nu)$ will decrease with increasing $T$, and we should get negative (positive) $\Delta R_{xx}$ for negative (positive) $B_{||}$.

In SI-Fig. 10a-c we have plotted the differential resistance, $dV_{xx}/dI$ vs. $I_{DC}$ with increasing temper-

15

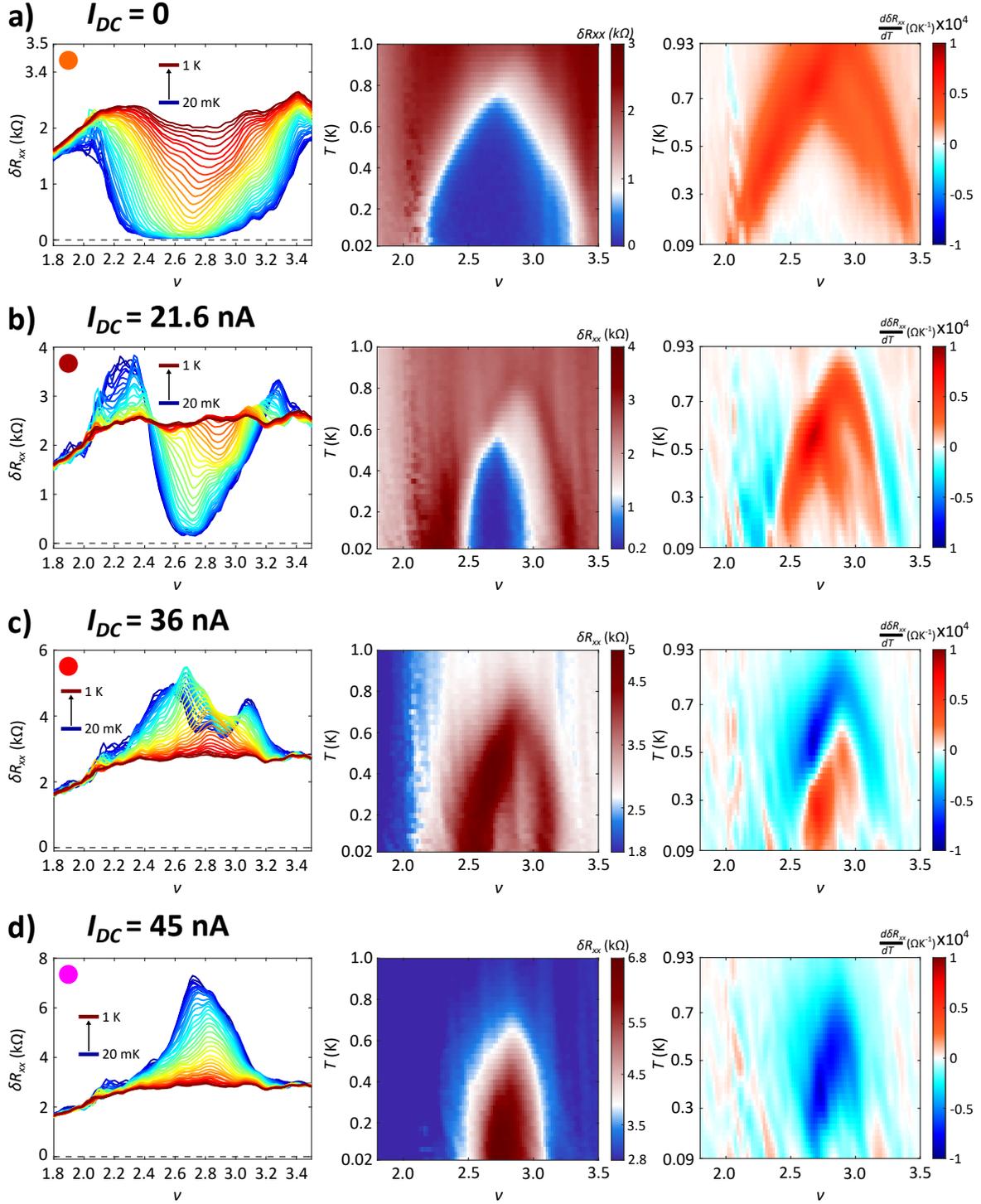

**SI-Fig. 11**: $\frac{dV_{xx}}{dI}$ **vs.** $\nu$ **with** $T$**:** (See Left to Right) Line plots of differential resistance, $dV_{xx}/dI (\equiv \delta R_{xx})$ vs. $\nu$ with increasing $T$, $\delta R_{xx}(\nu, T)$ 2−d colormap, and $(\nu, T)$ 2−d colormap of temperature derivative of differential resistance ($\frac{d\delta R_{xx}}{dT}$) at zero $D$ for $I_{DC}$ = **a)** 0, **b)** 21.60 nA, **c)** 36 nA, and **d)** 45 nA.

16

ature, $T$, for filling factors 2.2, 2.7, and 3.2. We observe that $dV_{xx}/dI$ increases with $T$ ($\frac{dR_{xx}}{dT}|_\nu > 0$), but at a finite value of $I_{DC}(\equiv I_s)$, $dV_{xx}/dI$ decreases with increasing $T$ ($\frac{dR_{xx}}{dT}|_\nu < 0$). Thus, as we see in SI-Fig.s 9b,d,f, for $I_{DC} > I_s$, increase in temperature with continuous $B_{||}$ sweep will lead to a 'downwards' BMR and switching in the sign of the rDMR.

To visualize the crossover between $\frac{dR_{xx}}{dT} > 0$ to $\frac{dR_{xx}}{dT} < 0$ (or vice versa) driven by $I_{DC}$, we have also plotted the temperature derivative of differential resistance for few selected values of $I_{DC}$. For the ease of notational representation, $dV_{xx}/dI$ has been replaced with $\delta R_{xx}$ in these plots. We see a robust superconducting dome for $I_{DC} = 0$ in SI-Fig. 11a (middle panel), where $\delta R_{xx}(\equiv R_{xx})$ increases with increasing $T$ in the filling range of $2 < \nu < 3.2$. $d\delta R_{xx}/dT$ is also positive for this region, as shown in SI-Fig. 11a (right panel). Normal 'upwards' hysteresis phase centred around $\nu \sim 2$ and $\sim 3$ reported in earlier sections and in main text all conform to this experimental condition. For $I_{DC} = 21.6$ (SI-Fig. 11b), the situation changes slightly. We have already seen that for $\nu = 2.2$ and $\nu = 3.2$, the value of switching current lies around $I_s \sim 14$ nA and $\sim 21-23$ nA. This means for $I_{DC} \sim 21$ nA, we see the superconducting dome being squeezed in the filling space as $\delta R_{xx}$ starts decreasing with increasing $T$ around the underdoped and overdoped region of the SC phase and we get $d\delta R_{xx}/dT < 0$ around these filling factors (SI-Fig. 11b (right panel)). With increasing $I_{DC}$ further, as we destroy the SC phase, the behavior of $\delta R_{xx}$ with $T$ is found to be completely reversed (SI-Fig. 11c,d with a negative value of $d\delta R_{xx}/dT$ taking up nearly whole of the filling phase space around $\nu \sim 2.4 - 3.2$ at $I_{DC} = 45$ nA (SI-Fig. 11d (right panel)) and we get a reversal of sign in rMR ($\Delta R_{xx}$) as obtained from the BMR hysteresis curve for these filling factors.

**SI- 7 : Relaxation time, $t_R$**

We can extract a relaxation time scale, $t_R$, of the transient decay of the measured $R_{xx}$ of near magic-angle TBG and local temperature as sensed by the Ruthenium oxide sensor (see Figure 5a,b of the main text) at different $B_{stop}$ as shown in SI-Fig. 12. '$t_R$' is defined as the time taken by the $R_{xx}$ (and $T_{Rox}$) to reach the steady state value from the value of $R_{xx}$ at $B_{stop}$. Interestingly, $t_R$ extracted for the TBG peaks around a $B_{stop}$ value of $-17$ mT, the value of $B_{||}$, where $R_{xx}^{FW}$, rises to a sharp peak ($\sim -B_{co}$). The difference in the $t_R$ values extracted for TBG and Rox sensors could be due to the different thermal masses of the two entities and/or variations in the cooling power of the fridge in different thermal cycles required to measure TBG and Rox temperature sensors separately. Though, the numbers are quite different for the extracted relaxation time ($t_R$) for the measured device and the sample stage (by virtue of the Rox sensor), the similarity in the behavior of the $t_R$ with $B_{stop}$, to our understanding, couldn't be a simple coincidence. Such evidences propel us to focus us on a possible extrinsic origin of the magnetoresistance hysteresis observed for the measured TBG device in the in-plane magnetic field ($B_{||}$).

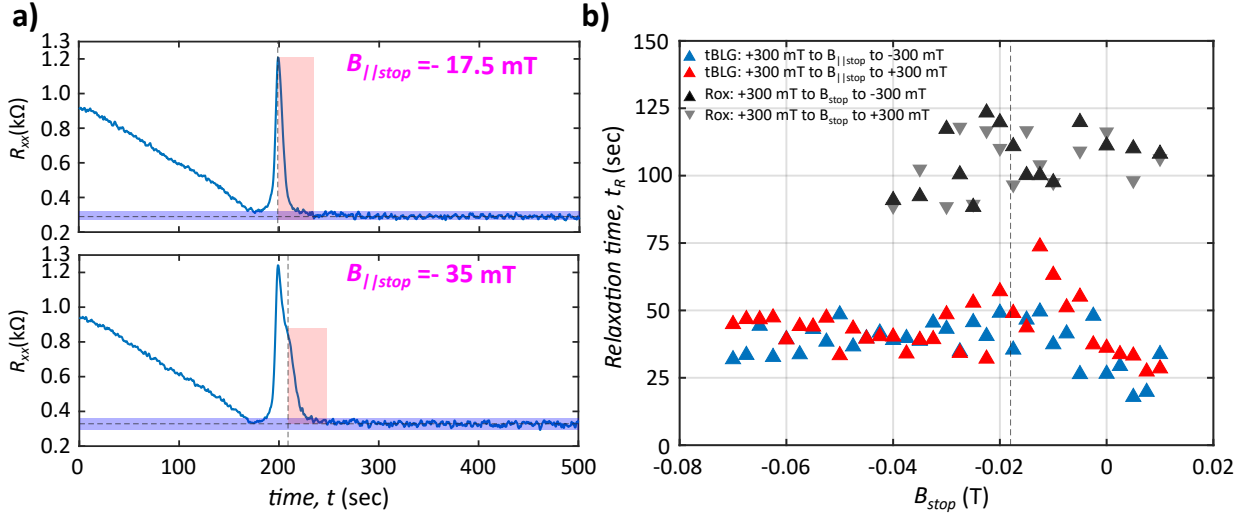

**SI-Fig. 12: Relaxation time: a)** Relaxation time, $t_R$, is defined as the time taken by the $R_{xx}$ (and $T_{Rox}$) to reach the steady state value from the value of $R_{xx}$ at $B_{stop}$. Extracting $t_R$ for near MATBG at $B_{\|stop} = -17.5$ mT (top panel) and $-35$ mT (bottom panel). **b)** $t_R$ vs. $B_{stop}$ for measured TBG and calibrated Rox sensor.

# Hysteresis at finite displacement fields

**SI- 8 : Controlling $n$ and $D$**

We have a dual-gated near magic-angle device where we have tuned the electrostatic doping and the applied perpendicular displacement field for our measurements. The metallic layer deposited on top of the top hBN and the SiO$_2$/Si act as the top gate and the global back gate, respectively. By applying top gate ($V_{tg}$) and back gate ($V_{bg}$) voltages on the device we can control the number density, $n$, and the vertical displacement field, $D$ [6–10].

**Controlling *n*:**
$$n = \frac{C_{bg}(V_{bg} - V_{bg,0}) + C_{tg}(V_{tg} - V_{tg,0})}{e} \qquad (2)$$

**Controlling *D*:**
$$D = \frac{C_{bg}(V_{bg} - V_{bg,0}) - C_{tg}(V_{tg} - V_{tg,0})}{2\epsilon_0} \qquad (3)$$

Here $C_{bg}$, $C_{tg}$, $V_{bg,0}$, $V_{bg,0}$, $e$ and $\epsilon_0$ are respectively the back gate capacitance per area, top gate capacitance per area, bottom gate Dirac point offset, top gate Dirac point offset, bare electronic charge, and free space

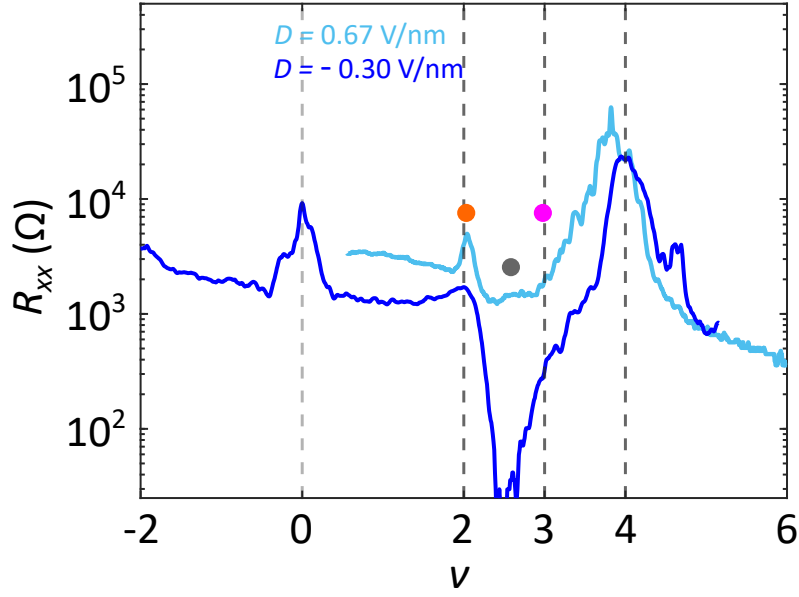

**SI-Fig. 13:** $R_{xx}$ **vs.** $\nu$ **at finite** $D$: $R_{xx}$ vs. filling factor, $\nu$, at $D = -0.30$ V/nm and $0.67$ V/nm values for $-2 < \nu < 6$ at a base temperature of 20 mK in absence of any magnetic field. The three solid circles correspond to filling ($\nu$) values: 2.1 (●), 2.6 (●), and 3.0 (●) respectively.

(vacuum) permittivity. The thicknesses of the encapsulating top and bottom hBN layers are $\sim 28$ nm and $\sim 30$ nm, respectively.

Measured longitudinal resistance, $R_{xx}$, in SI-Fig. 13 for different $D$ values show the tuning behavior of the superconducting phase with $D$ [11] between filling factor, $\nu(\equiv 4n/n_s) \approx 2-3$.

## SI- 9 : Downwards BMR at finite $D$

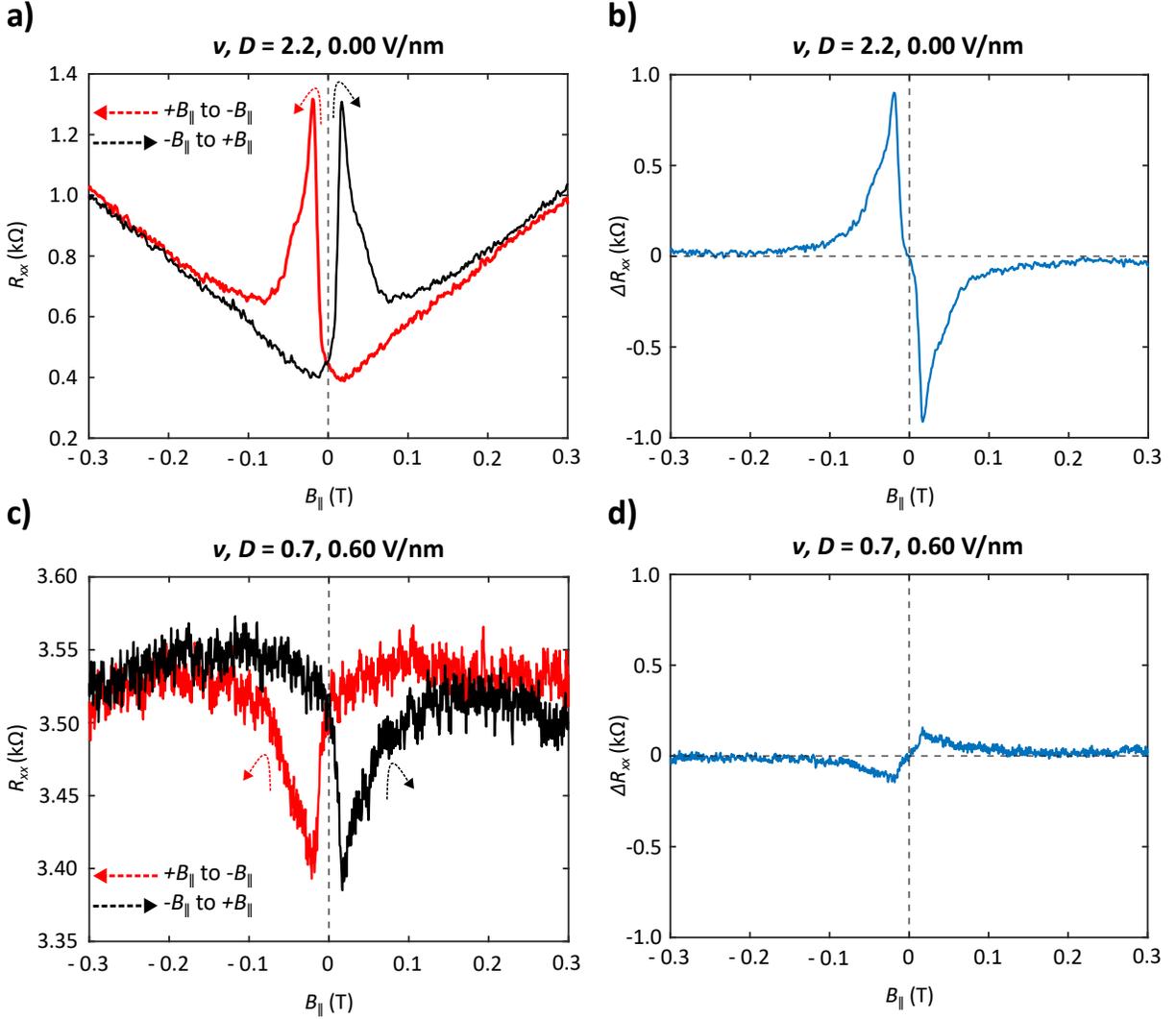

**SI-Fig. 14: Comparing 'upwards' and 'downwards' BMR in near magic-angle TBG: a)** BMR hysteresis curve measured for $(\nu, D) = 2.2, 0.00$ V/nm. We have defined the red $R_{xx}$ trace with $B_{||}$ $(+300$ mT $\rightarrow -300$ mT) as $R_{xx}^{FW}$. Similarly, the **black** $R_{xx}$ trace with $B_{||}$ $(-300$ mT $\rightarrow +300$ mT) sweep has been defined as $R_{xx}^{BW}$. We see the appearance of a sharp peak in $R_{xx}$ at $\sim -18$ mT $(+18$ mT) for the forward (backward) sweep directions. **b)** $\Delta R_{xx}$ for $(\nu, D) = 2.2, 0.00$ V/nm. **c)** BMR hysteresis curve measured at $(\nu, D) = 0.7, 0.60$ V/nm. Sharp $R_{xx}$ dips appear around $\sim -17.5$ mT $(+17.5$ mT) mT for the forward (backward) sweep directions. **d)** $\Delta R_{xx}$ for $(\nu, D) = 0.7, 0.60$ V/nm. We notice a reversal in sign of $\Delta R_{xx}$ for a given $B_{||}$ for two different choice of $(\nu, D)$ in **a)** and **c)** at an operational base temperature of $\sim 65-70$ mK.

20

## SI- 10 : Hysteresis around the superconducting pocket at finite $D$

For $(\nu, D) = 2.6, -0.30$ V/nm when the SC phase is relatively weaker but not suppressed enough, signatures of $R_{xx}$ hysteresis are absent (SI-Fig. 15b). As soon as we move to the +ve $D$ side, which has a more stronger suppression effect on the SC, at $(\nu, D) = 2.6, 0.60$ V/nm [11], we recover the familiar shape of BMR hysteresis response with a finite value of reduced magnetoresistance (rMR), $|\Delta R_{xx}|$ (SI-Fig. 15e). It is striking to note that for any given $|D|$, the boundaries of the SC phase on the filling axis, *i.e.*, $\nu \sim 2.2$ and $\nu \sim 3$, we always have a $R_{xx}$ hysteresis with $B_{||}$ sweep and a finite rMR, $|\Delta R_{xx}|$. This observation can be seen more easily from the $\Delta R_{xx}(\nu, B_{||})$ 2−d colormap in SI-Fig. 16a,b,c for three different values of $D = -0.30$ V/nm, $0.00$ V/nm, and $0.60$ V/nm. For $D = -0.30$ V/nm and $0.00$ V/nm we see two separate hysteresis 'islands' around $\nu \sim 2$ and $\nu \sim 3$ (SI-Fig. 16a,b) with similar sign $(+/-)$ and magnitude of $\Delta R_{xx}$ with $\mp B_{||}$. These two islands are observed to have merged in SI-Fig. 16c by a weak inter-connecting hysteresis region at $D = 0.60$ V/nm. We also see that between $\nu \sim 0.5 - 1.3$ for $D = 0.00$ V/nm and $0.60$ V/nm, there is a reversal of the sign of the MR with $|B_{||}|$, highlighted by a dashed box in SI-Fig. 16b,c, the implications of which will be discussed in later sections.

## SI- 11 : Effect of temperature of the MC plate with $B_{||}$ sweep at finite $D$

Similar to SI- 3, We have tried to see if a change of $T_{MC}$ with a continuous $B_{||}$ sweep has any effect on the observed nature of BMR response and $\Delta R_{xx}$ magnitude at finite $D$ magnitudes. Similarly $R_{xx}^{FW}$ and $R_{xx}^{BW}$, $T_{MC}^{FW}$ and $T_{MC}^{BW}$ are plotted in SI-Fig. 17 for different values of $(\nu, D)$. As shown in the bottom panels of SI-Fig. 17a-f, the $T_{MC}$ profile for the 'forward' and 'backward' sweep are always similar for any choice of $(\nu, D)$ irrespective of the nature of BMR (hysteresis or no hysteresis). Here also, the profile of the increase in temperature is monotonic which does not correlate to the non-monotonic increase or decrease of $R_{xx}$ vs. $B_{||}$ curves.

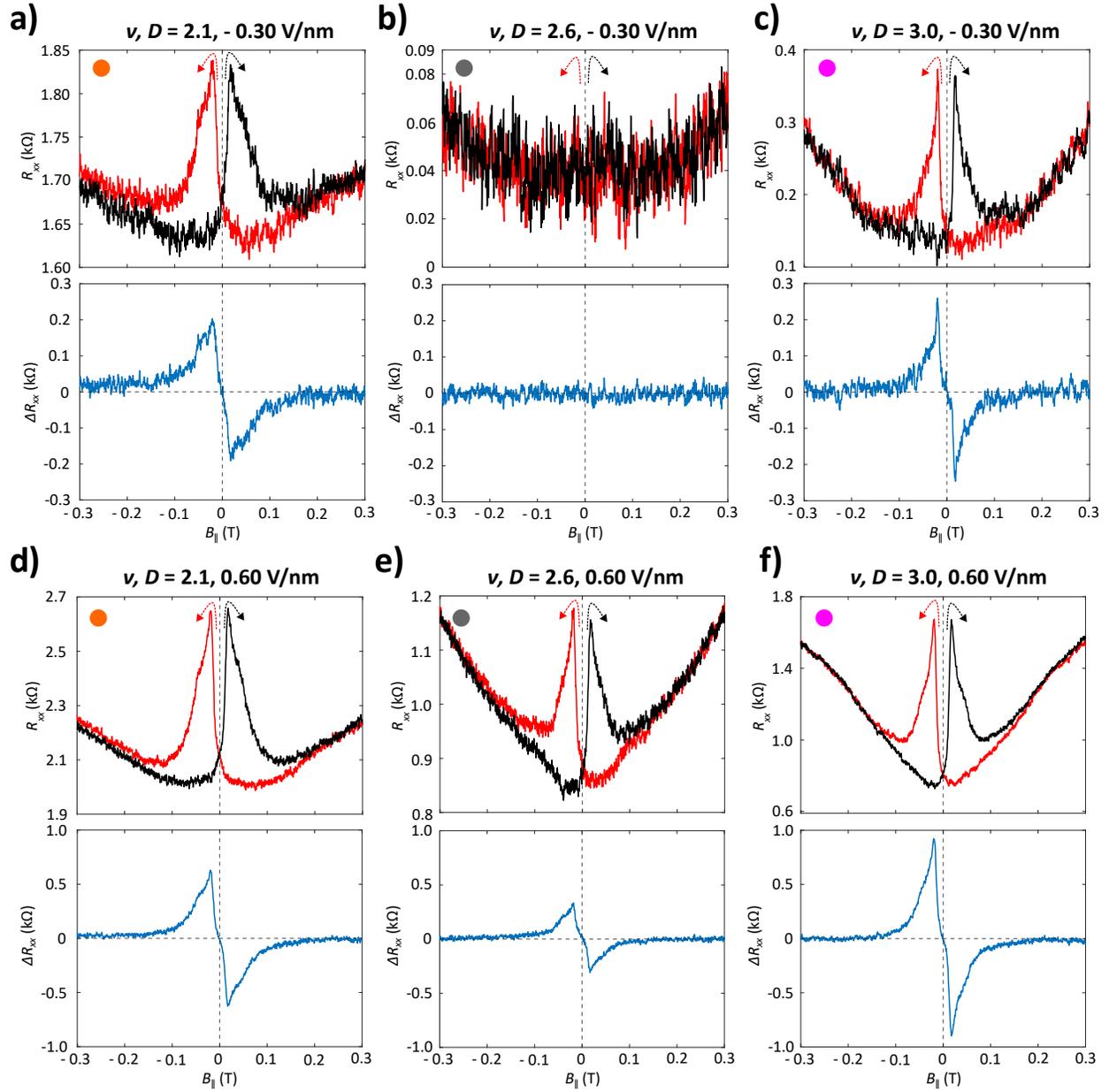

**SI-Fig. 15: Hysteresis in $R_{xx}$ in the $\nu - D$ phase space:** BMR hysteresis traces ($R_{xx}^{FW}$ and $R_{xx}^{BW}$) (top panel) and calculated $\Delta R_{xx}$ (bottom panel) within the $B_{\parallel}$ sweep range of $\pm 300$ mT for **a)** $(\nu, D) = 2.1, -0.30$ V/nm, **b)** $(\nu, D) = 2.6, -0.30$ V/nm, **c)** $(\nu, D) = 3.0, -0.30$ V/nm, **d)** $(\nu, D) = 2.1, 0.60$ V/nm, **e)** $(\nu, D) = 2.6, 0.60$ V/nm, **f)** $(\nu, D) = 3.0, 0.60$ V/nm. The $B_{\parallel}$ sweep rate is: 100 mT/min.

22

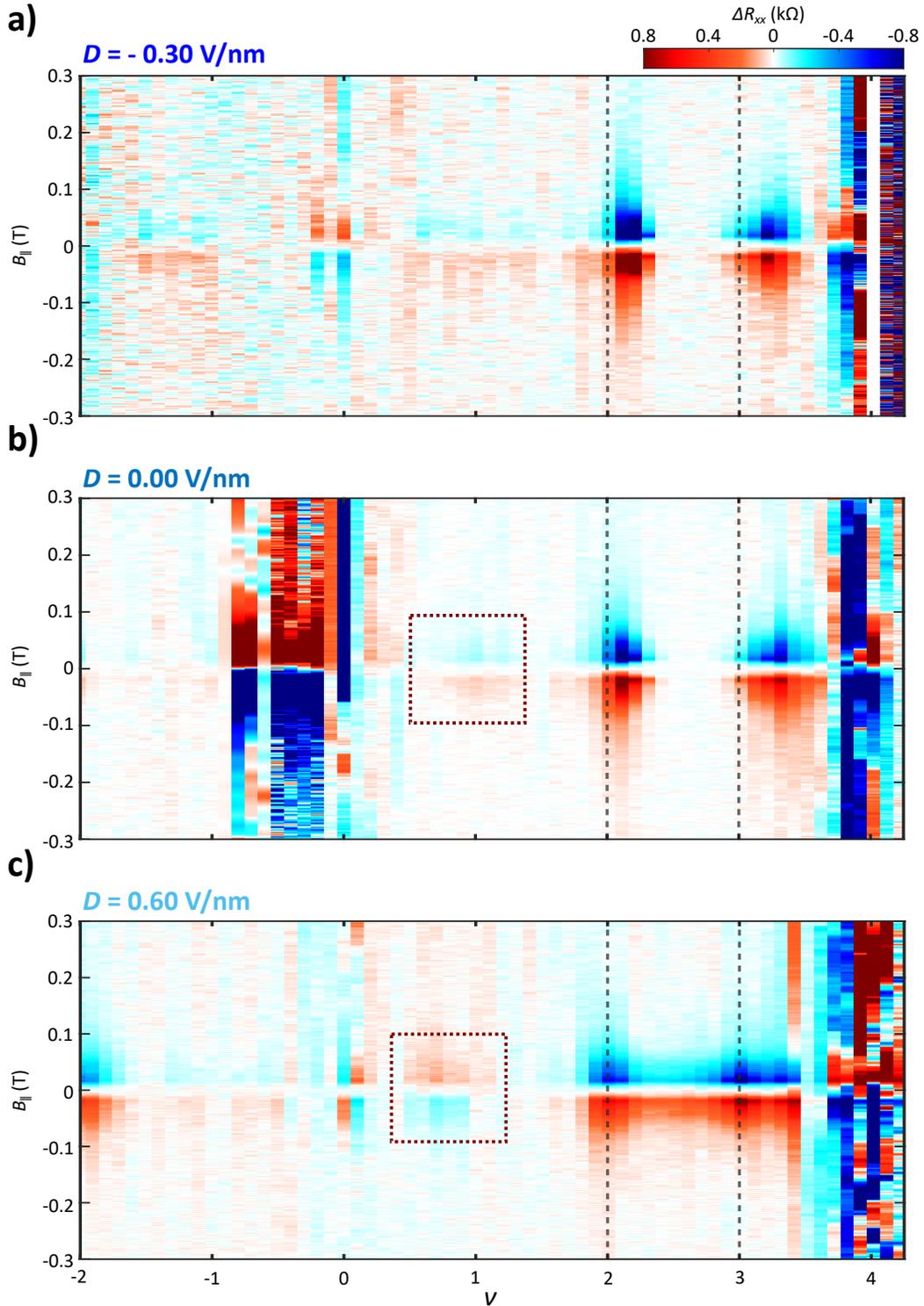

**SI-Fig. 16: rMR map in the $\nu - B_{||}$ phase space with $D$:** $\Delta R_{xx}(\nu, B_{||})$ 2−d colormap for $D =$ **a)** $-0.30$ V/nm, **b)** $0.00$ V/nm, and **c)** $0.60$ V/nm at an effective sample/base temperature of 70 mK. We also see a reversal of sign in $\Delta R_{xx}$ between $D = 0.00$ V/nm and $0.60$ V/nm around $\nu \sim 0.3 - 1.3$, outlined in the dotted square, compared to the sign of $\Delta R_{xx}$ in the hysteresis region of $\nu \sim 2 - 3$.

23

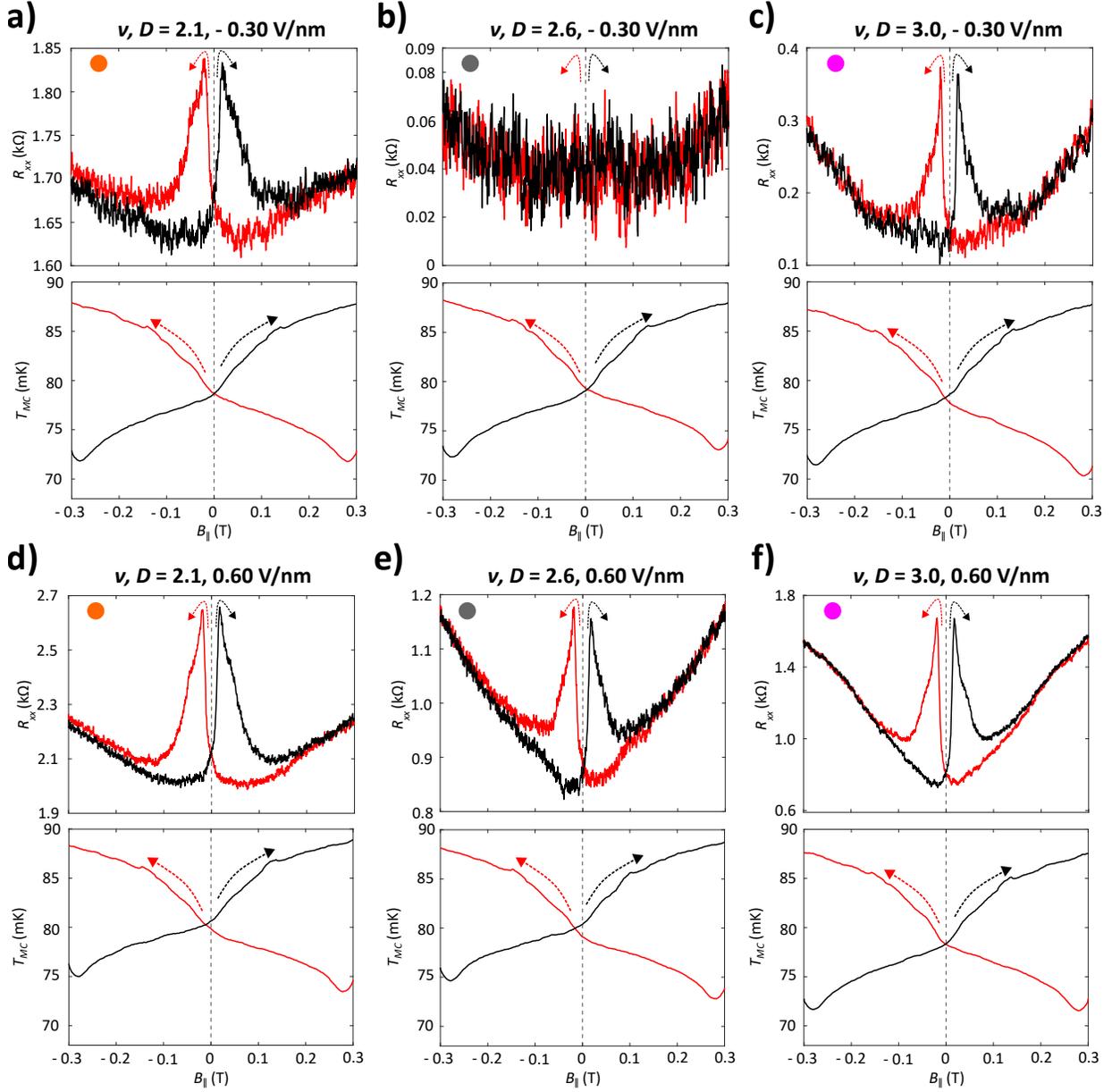

SI-Fig. 17: $R_{xx}$ **and measured** $T_{MC}$ **profile with forward and backward** $B_{||}$ **sweep at finite** $D$: BMR hysteresis traces ($R_{xx}^{FW}$ and $R_{xx}^{BW}$) (top panel) and mixing chamber plate/bath temperature profile ($T_{MC}^{FW}$ and $T_{MC}^{BW}$) (bottom panel) within the $B_{||}$ sweep range of $\pm 300$ mT for **a)** $(\nu, D) = 2.1, -0.30$ V/nm, **b)** $(\nu, D) = 2.6, -0.30$ V/nm, **c)** $(\nu, D) = 3.0, -0.30$ V/nm, **d)** $(\nu, D) = 2.1, 0.60$ V/nm, **e)** $(\nu, D) = 2.6, 0.60$ V/nm, **f)** $(\nu, D) = 3.0, 0.60$ V/nm. The $B_{||}$ sweep rate is: 100 mT/min. At the start of the hysteresis loop $T_{MC}^{FW}$ rises continuously from $\sim 72$ mK to $\sim 90$ mK as $B_{||}$ is changed continuously from $+300$ mT to $-300$ mT. We add a delay of 2 minutes before we switch back to the sweep direction. Due to this $T_{MC}$ decreases down to $\sim 75$ mK, closer to the initial temperature at the beginning of the hysteresis loop. The rise of $T_{MC}^{BW}$ is similar with the opposite direction of $B_{||}$ sweep. We see that even if there is a temperature difference of $\Delta T_{MC} \sim 10$ mK between the two sweep directions for $|B_{||}| \sim 150 - 300$ mT, $R_{xx}^{FW} \approx R_{xx}^{BW}$ for all values of $(\nu, D)$ similar to as shown for zero $D$ in SI-Fig. 3

## SI- 12 : Evolution of hysteresis phase with temperature at $D > 0$

$D = 0.60$ V/nm has a single hysteresis phase spanning over the filling range $\nu \sim 1.9 - 3.4$ as seen from SI-Fig. 16c. Similar to the observations in Figure 2 in the main text, near the filling factors where we see stronger hysteresis, *i.e.*, $\nu = 2.1, 3.0$, we see the similar weakening of the $\Delta R_{xx}$ with increasing temperature as shown in SI-Fig. 18a,c. The hysteresis phase around $(\nu, D) = 3.0, 0.60$ V/nm is observed to be more stronger compared to the one at $(\nu, D) = 2.1, 0.60$ V/nm with a higher temperature value required to achieve the collapse of rMR ($\Delta R_{xx} \approx 0$). In comparison, the weaker hysteresis phase appearing around $\nu = 2.5$ at the base operational temperature is seen to persist even upto 1 K with little to no observational weakening with increasing $T$.

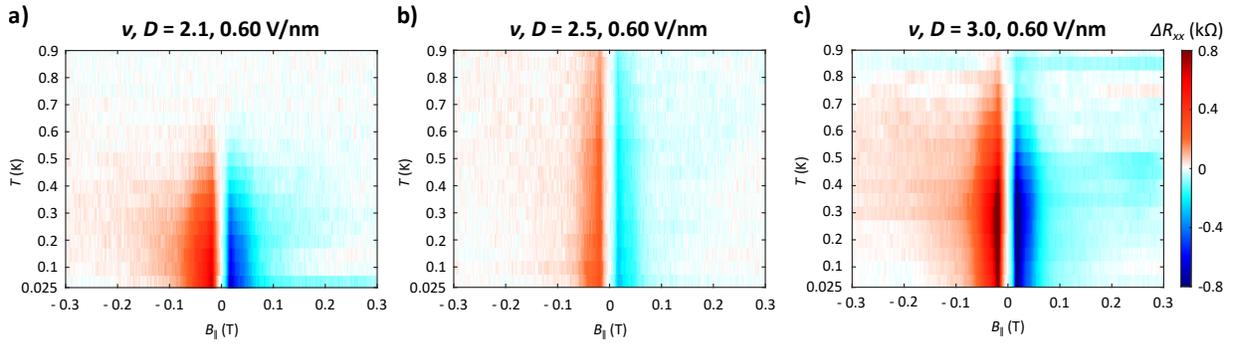

**SI-Fig. 18: Hysteresis with $T$ at $D = 0.60$ V/nm:** $\Delta R_{xx}(B_{\parallel}, T)$ 2−d colormap at $D = 0.60$ V/nm for $\nu = $ **a)** 2.1, **b)** 2.5, and **c)** 3.0.

## SI- 13 : Transient effect in $R_{xx}(B_{\parallel})$ at higher $D$

We have already discussed the observation of the transient effect in the hysteresis phase which appears at zero $D$. Now the next question is, is it also present for the hysteresis phase observed at higher $D$? The answer is 'yes'. SI-Fig. 19a shows a comparison of BMR curves and magnitude of MR, $\Delta R_{xx}$, between $D = 0.00$ V/nm and 0.60 V/nm at $\nu = 2.2$. SI-Fig. 19b shows the comparison for the transient sweep scheme:+300 mT$\rightarrow B_{\parallel stop} \rightarrow +300$ mT for these two different $D$ fields for $B_{\parallel stop} = -50$ mT. We observe that even though there is a stark difference in the values of $R_{xx}^{FW}$, $R_{xx}^{BW}$, and $|\Delta R_{xx}|$ between two different $D$ values, the nature of transient $R_{xx}$ response remains essentially similar for both the $D$ fields.

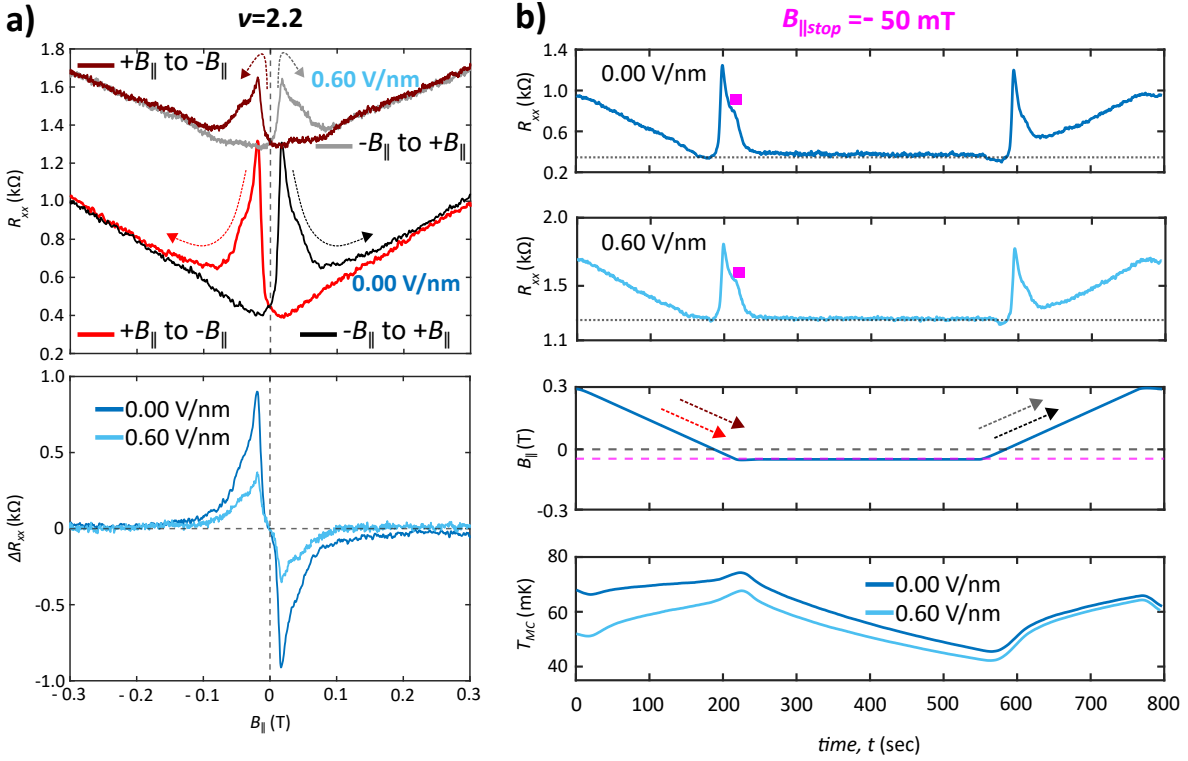

**SI-Fig. 19: Transient effect in hysteresis at higher $D$: a)** BMR hysteresis traces ($R_{xx}^{FW}$ and $R_{xx}^{BW}$) (top panel) and calculated $\Delta R_{xx}$ (bottom panel) within the $B_\parallel$ sweep range of $\pm 300$ mT for $D = 0.00$ V/nm and $0.60$ V/nm at $\nu = 2.2$. **b)** $R_{xx}$, $B_\parallel$ and $T_{MC}$ vs. time, $t$, plotted for $B_{\parallel stop} = -50$ mT. $R_{xx}$ transiently relaxes to a steady state value after the magnetic field is stopped ramping and held at $-50$ mT just at the beginning of stop interval of 300 seconds for both the $D$ values.

## SI- 14 : Ubiquitous nature of the transient effect

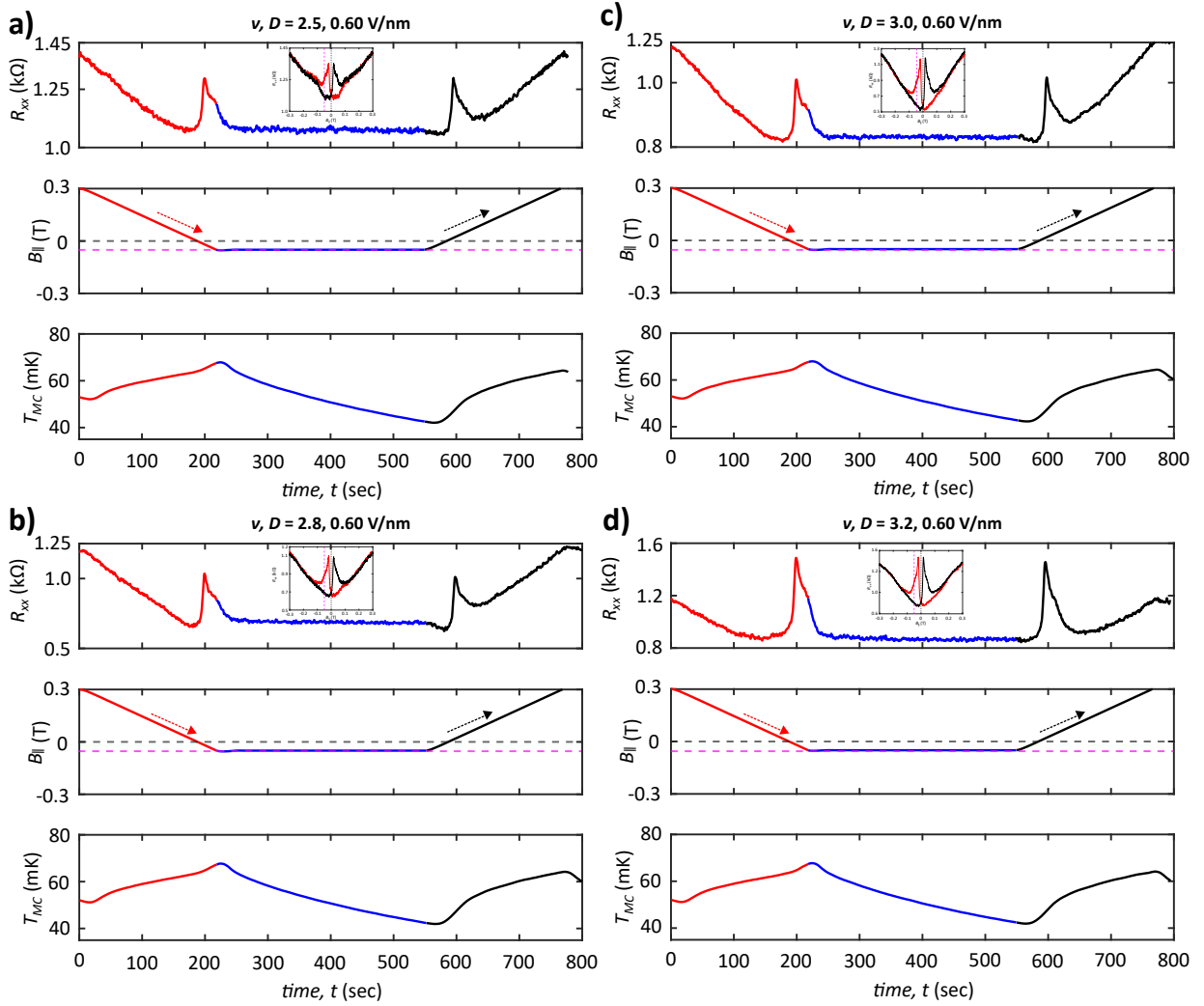

**SI-Fig. 20**: **Transient effect observed in $R_{xx}$ at different filling factors at finite $D$**: $R_{xx}$, $B_{||}$, and $T_{MC}$ for the backward (SI- 5) transient sweep with time, $t$, plotted in three panels for $B_{||stop} = -50$ mT (horizontal dashed line in the middle panels in magenta) near a displacement field of $D = 0.60$ V/nm and at $\nu$ values of **a)** 2.5, **b)** 2.8, **c)** 3.0, and **d)** 3.2. We observe for any filling where we have BMR hysteresis (insets in top panels of each figures), a transient effect is ubiquitously present. Red arrow (↘) in the middle panels indicates to the magnetic sweep direction in $+300$ mT to $B_{||stop}$ and the **Black** arrow (↗) indicate to the magnetic sweep direction in $B_{||stop}$ to $+300$ mT.

## SI- 15 : Current driven ($I_{DC}$) switching in rMR/rDMR at higher $D$

In SI-Fig. 21 we have plotted differential resistance, $dV_{xx}/dI$, as a function of D.C. bias current, $I_{DC}$, for different filling factors $\nu = 2.2, 2.7$, and $3.2$.

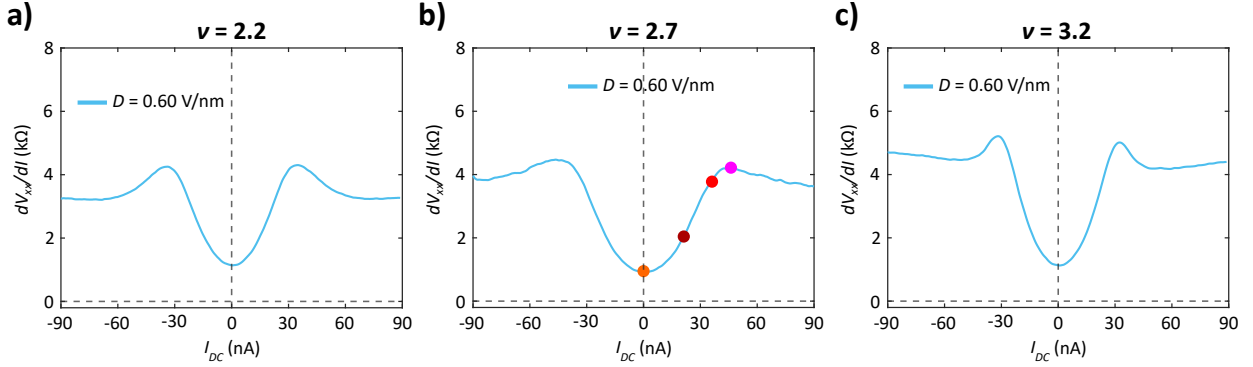

**SI-Fig.** 21: $dV_{xx}/dI$ vs. $I_{DC}$: Differential resistance is plotted as a function of applied $I_{DC}$ for 0.60 V/nm at filling factors of $\nu =$ **a)** 2.2, **b)** 2.7, and **c)** 3.2. Solid circles on the $dV_{xx}/dI$ curve of **(b)** are at different $I_{DC}$ values: 0 (●), 21.60 nA (●), 36 nA (●), and 45 nA (●).

### Hysteresis at finite $D$ in presence of applied $I_{DC}$ under $B_\parallel$

SI-Fig.s 22a,c,e show how differential resistance, $dV_{xx}/dI$, changes with increasing $B_\parallel$ upto $+300$ mT for $|I_{DC}| \geq 0$ for three representative filling factors at $D = 0.60$ V/nm. As seen in SI- 15 for zero $D$, with increasing $B_\parallel$ at $D = 0.60$ V/nm, $dV_{xx}/dI$ is also seen to increase upto a certain value of $I_{DC}$ before the behavior is switched (defined as $I_s$), *i.e.*, $dV_{xx}/dI$ decreases with increasing $B_\parallel$. We can perform the hysteresis measurement at a fixed value of applied $I_{DC}$ by measuring $dV_{xx}^{FW}/dI$ and $dV_{xx}^{BW}/dI$ with forward and backward $B_\parallel$ sweep and calculate the reduced differential MR or rDMR defined in equation (1). Similar to in SI-Fig. 9, if the applied $I_{DC}$ value is smaller or greater than $I_s$ for a choice of $\nu$ at $D = 0.60$ V/nm, we get a reversal of sign in rDMR where we cross over from positive (negative) to negative (positive) rDMR as we change applied D.C. bias current from $I_{DC} < I_s$ to $I_{DC} > I_s$ as demonstrated in $\Delta \frac{dV_{xx}}{dI}(B_\parallel, I_{DC})$ $2-$d colormaps in SI-Fig.s 22b,d,f before the collapse of rDMR at a higher bias current ($I_{DC} > 60$ nA).

In SI-Fig.s 23a,b,c, we have plotted the differential resistance, $dV_{xx}/dI$ vs. $I_{DC}$ with increasing temperature, $T$, for $\nu$ values of 2.2, 2.8, and 3.2 at $D = 0.60$ V/nm. We observe that $dV_{xx}/dI$ increases with $T$ ($\frac{dR_{xx}}{dT}|_\nu > 0$), but at a finite value of $I_{DC}(\equiv I_s)$, $dV_{xx}/dI$ decreases with increasing $T$ ($\frac{dR_{xx}}{dT}|_\nu < 0$). Thus, as we see in SI-Fig.s 22b,d,f, for $I_{DC} > I_s$, increase in temperature with $B_\parallel$ sweep will lead to a 'downwards' BMR and switching in the sign of the rDMR.

28

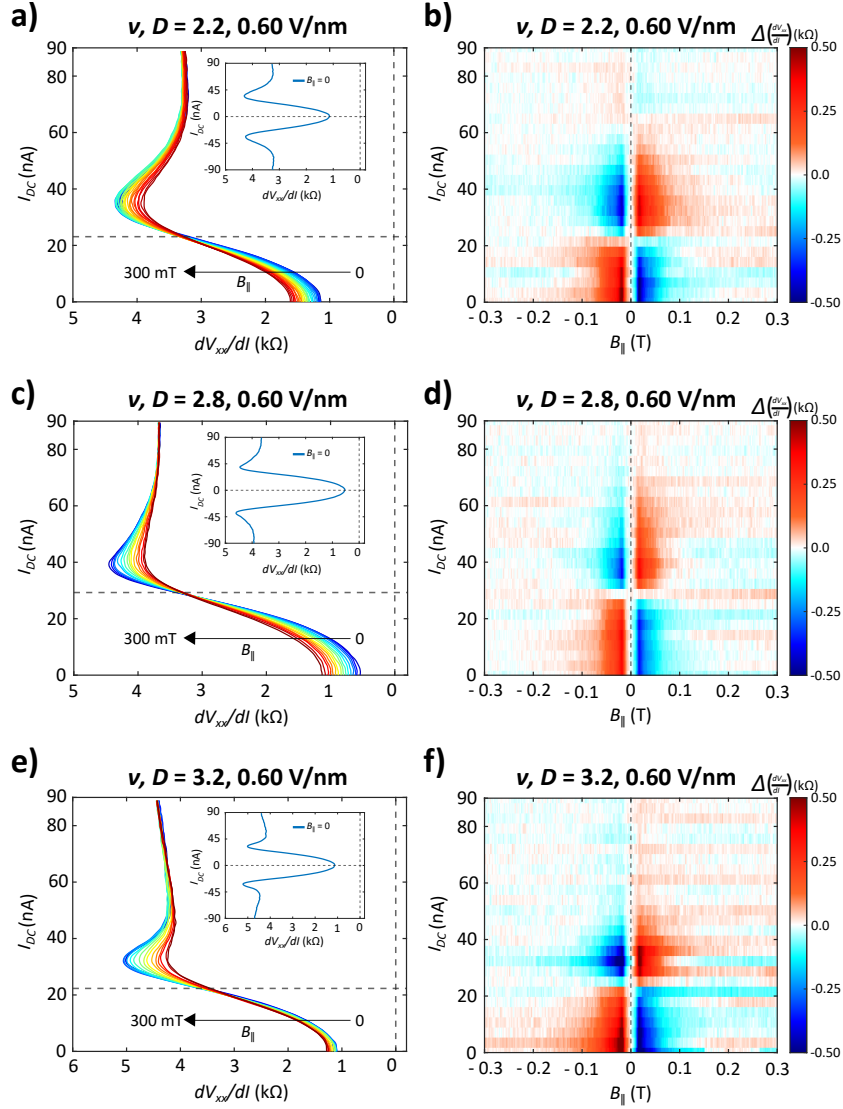

**SI-Fig. 22: Sign reversal of rDMR at $D = 0.60$ V/nm with $I_{DC}$:** $\frac{dV_{xx}}{dI}$ plotted as a function of $I_{DC}$, with increasing in-plane magnetic field from 0 to 300 mT for $D = 0.60$ V/nm at $\nu$ values of **a)** 2.2, **c)** 2.8, and **e)** 3.2. $dV_{xx}/dI$ traces have been shown only for positive values of $I_{DC}$. The behavior is symmetric across zero current for $I_{DC} < 0$. Inset shows the $dV_{xx}/dI$ vs. $I_{DC}$ for the corresponding filling factors at $B_{||} = 0$. We observe that, initially $dV_{xx}/dI$ increases with increasing $B_{||}$ but the behavior reverses ($dV_{xx}/dI$ decreases with increasing $B_{||}$) at a sufficiently higher value of $I_{DC}$ defined as $I_s$. The $I_s$ values for the three filling factors are $\sim 24.5$ nA ($\nu = 2.2$), $\sim 29$ nA ($\nu = 2.8$), and $\sim 21$ nA ($\nu = 3.2$). The reversal in the nature $dV_{xx}/dI$ with applied $B_{||}$ below and above $I_s$ is also reflected in the differential magnetoresistance (rDMR). Below $I_s$, the BMR response resembles the nature as shown in SI-Fig. 14a, and above $I_s$, the BMR response is flipped like in SI-Fig. 14c. As a result, we also see a reversal of sign in rDMR for a given $B_{||}$ as seen in the 2−d colormap of $\Delta \frac{dV_{xx}}{dI}(B_{||}, I_{DC})$ for $\nu =$ **b)** 2.2, **d)** 2.8, and **f)** 3.2. The values of $I_{DC}$ where rDMR shows the sign reversal in the colormaps match very closely to the $I_s$ values for each $\nu$ found from the line plots of $dV_{xx}/dI$ vs. $I_{DC}$ with increasing $B_{||}$.

29

To visualize the crossover between $\frac{dR_{xx}}{dT} > 0$ to $\frac{dR_{xx}}{dT} < 0$ (or vice versa) driven by $I_{DC}$ at finite $D$, we again plotted the temperature derivative of differential resistance for few selected values of $I_{DC}$. For the ease of notational representation of $dV_{xx}/dI$, it has been replaced with $\delta R_{xx}$ in these plots. In SI-Fig. 24a (middle panel) $\delta R_{xx} (\equiv R_{xx})$ increases with increasing $T$ in the filling range of $2 < \nu < 3.2$. $d\delta R_{xx}/dT$ is also positive for this region as shown in SI-Fig. 24a (right panel). Normal 'upwards' hysteresis phase centred around $\nu \sim 2$ and $\sim 3$ at $D = 0.60$ V/nm conform to this experimental condition. The situation changes when $I_{DC} \sim 21$ nA, $\delta R_{xx}$ starts decreasing with increasing $T$ around $\nu \sim 2$ and $\sim 3$, and we get $d\delta R_{xx}/dT < 0$ around these filling factors (SI-Fig. 24b (right panel)). With increasing $I_{DC}$ further, the behavior of $\delta R_{xx}$ with $T$ is found to be completely reversed (SI-Fig. 24c,d) with a negative value of $d\delta R_{xx}/dT$ taking up nearly whole of the filling phase space around $\nu \sim 2.4 - 3.2$ at $I_{DC} = 45$ nA (SI-Fig. 24d (right panel))

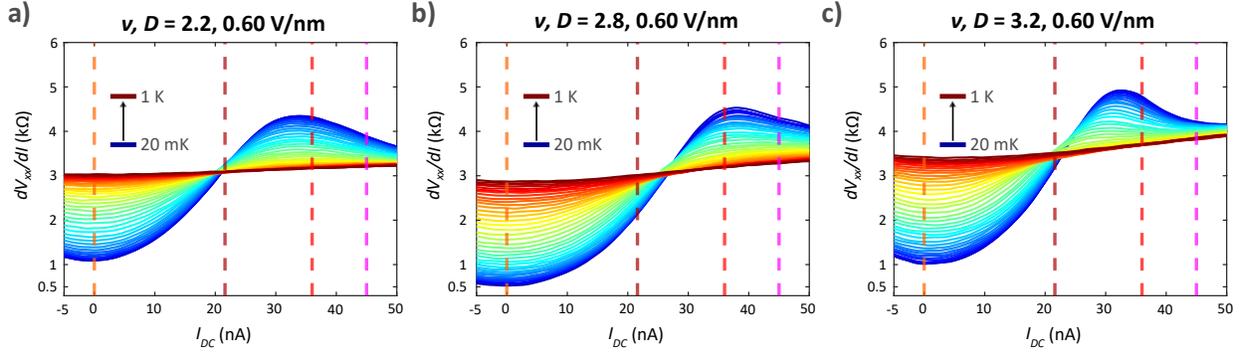

**SI-Fig. 23:** $\frac{dV_{xx}}{dI}$ **vs.** $I_{DC}$ **at** $D = 0.60$ **V/nm with** $T$**:** $\frac{dV_{xx}}{dI}$ plotted as a function of applied $I_{DC}$ with increasing $T$ from 20 mK to 1 K for $D = 0.60$ V/nm at $\nu$ values of **a)** 2.2, **b)** 2.8, and **c)** 3.2. Dashed vertical lines are for $I_{DC}$ values of 0, 21.60 nA, 36 nA, and 45 nA. $dV_{xx}/dI$ increases with increasing $T$ but the behavior reverses ($dV_{xx}/dI$ decreases with increasing $T$) at a sufficiently higher value of $I_{DC}$ defined as $I_s$. The $I_s$ values for the three filling factors are $\sim 21$ nA ($\nu = 2.2$), $\sim 25$ nA ($\nu = 2.8$), and $\sim 22$ nA ($\nu = 3.2$).

30

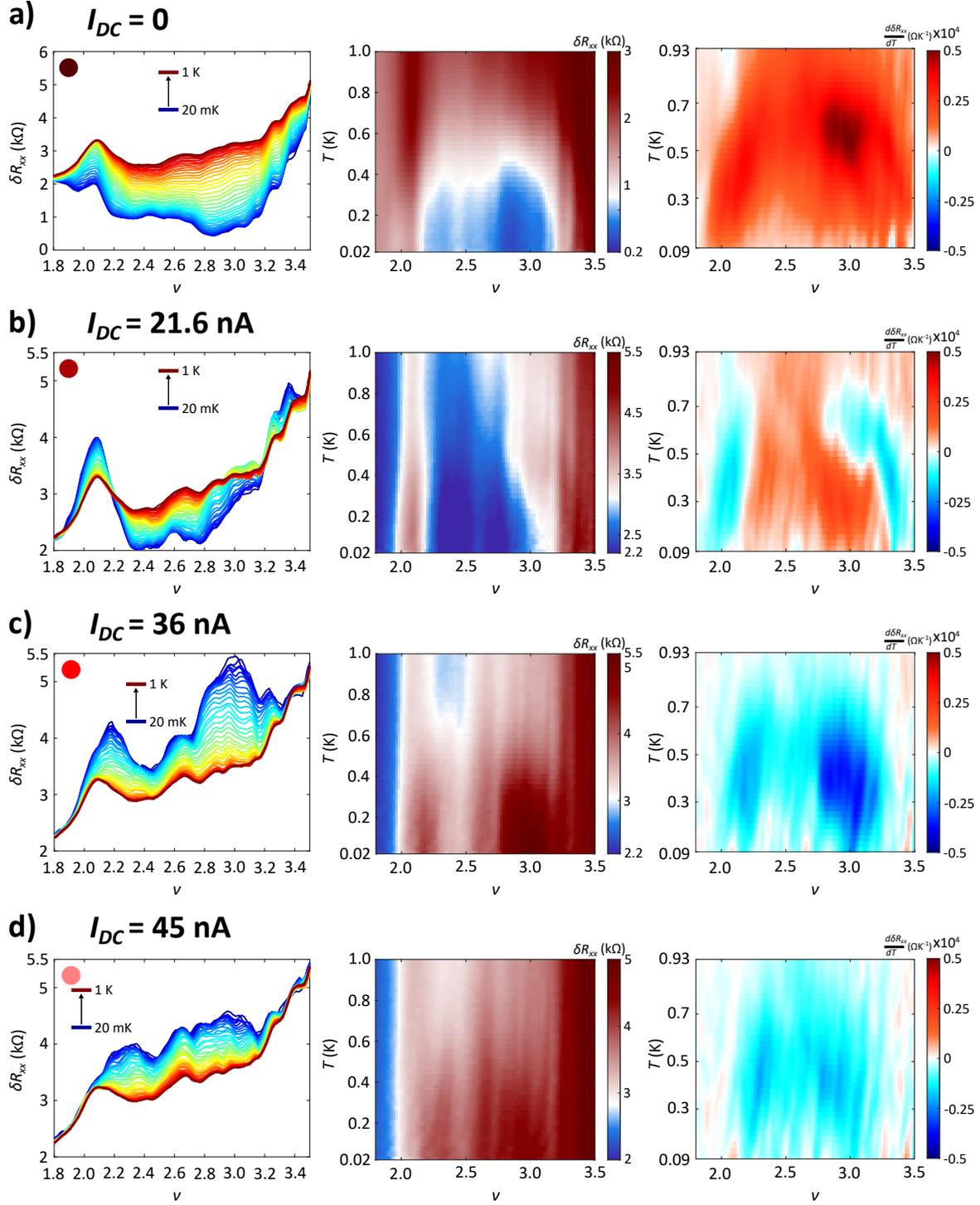

**SI-Fig. 24:** $\frac{dV_{xx}}{dI}$ **vs.** $\nu$ **with** $T$ **at** $D = 0.60$ **V/nm:** (See Left to Right) Line plots of differential resistance, $dV_{xx}/dI (\equiv \delta R_{xx})$ vs. $\nu$ with increasing $T$, $\delta R_{xx}(\nu, T)$ 2−d colormap, and $(\nu, T)$ 2−d colormap of temperature derivative of differential resistance ($\frac{d\delta R_{xx}}{dT}$) at $D = 0.60$ V/nm for $I_{DC} =$ **a)** 0, **b)** 21.60 nA, **c)** 36 nA, **d)** 45 nA.

31

## SI- 16 : Sign reversal of rMR around $\nu \sim 0.75$ between zero and higher $D$

We saw a reversal of sign in the observed $\Delta R_{xx}$ centred around the filling factor of $\nu \sim 0.75$ between $D = 0.00$ V/nm and $D = 0.60$ V/nm in SI-Fig. 16b,c. Following our discussion in SI- 6 and SI- 15 we are now able to recognize that this could originate from a crossover from a $dR/dT > 0$ to a $dR/dT < 0$ state with applied $D$. Depending on that we get either one of the two BMR hysteresis loop as shown in SI-Fig. 14a,c and a reversal of sign in MR (for a given $|B_{||}|$). Temperature map of $R_{xx}$ around filling $\nu \sim 0.75$ for zero $D$ and $D > 0$ is shown in the top panels of SI-Fig. 25a,b. The bottom panels show how $dR_{xx}/dT$ changes with increasing $T$ in the filling range $\nu \sim 0.5 - 1.5$ for two different regimes of displacement field strength. It is evident that at $D = 0.00$ V/nm resistive state at $\nu \sim 0.75$ has $dR/dT > 0$, which transitions to a state with $dR/dT < 0$ at higher $D$, essentially facilitating the observed sign reversal in MR with $B_{||}$ sweep.

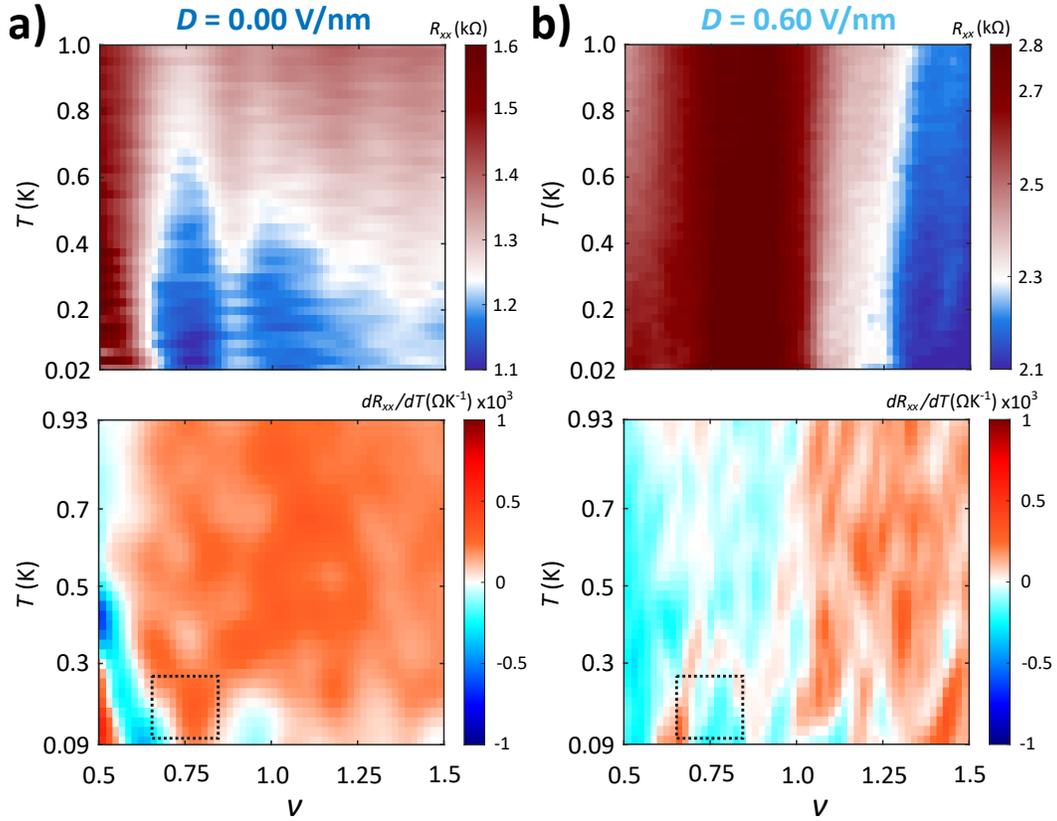

**SI-Fig. 25: Sign reversal of rMR around $\nu \sim 0.75$: a)** $R_{xx}(\nu, T)$ 2−d colormap (top panel), and $\frac{dR_{xx}}{dT}(\nu, T)$ 2−d colormap (bottom panel) of the same for $D = 0.00$ V/nm between filling factors $\nu \sim 0.5 - 1.5$. **b)** $R_{xx}(\nu, T)$ 2−d colormap (top panel), and $\frac{dR_{xx}}{dT}(\nu, T)$ 2−d colormap (bottom panel) of the same for $D = 0.60$ V/nm between filling factors $\nu \sim 0.5 - 1.5$. The data have been taken for $B_{||} = 0$.

## SI- 17 : Hysteresis with out-of-plane magnetic field ($B_\perp$)

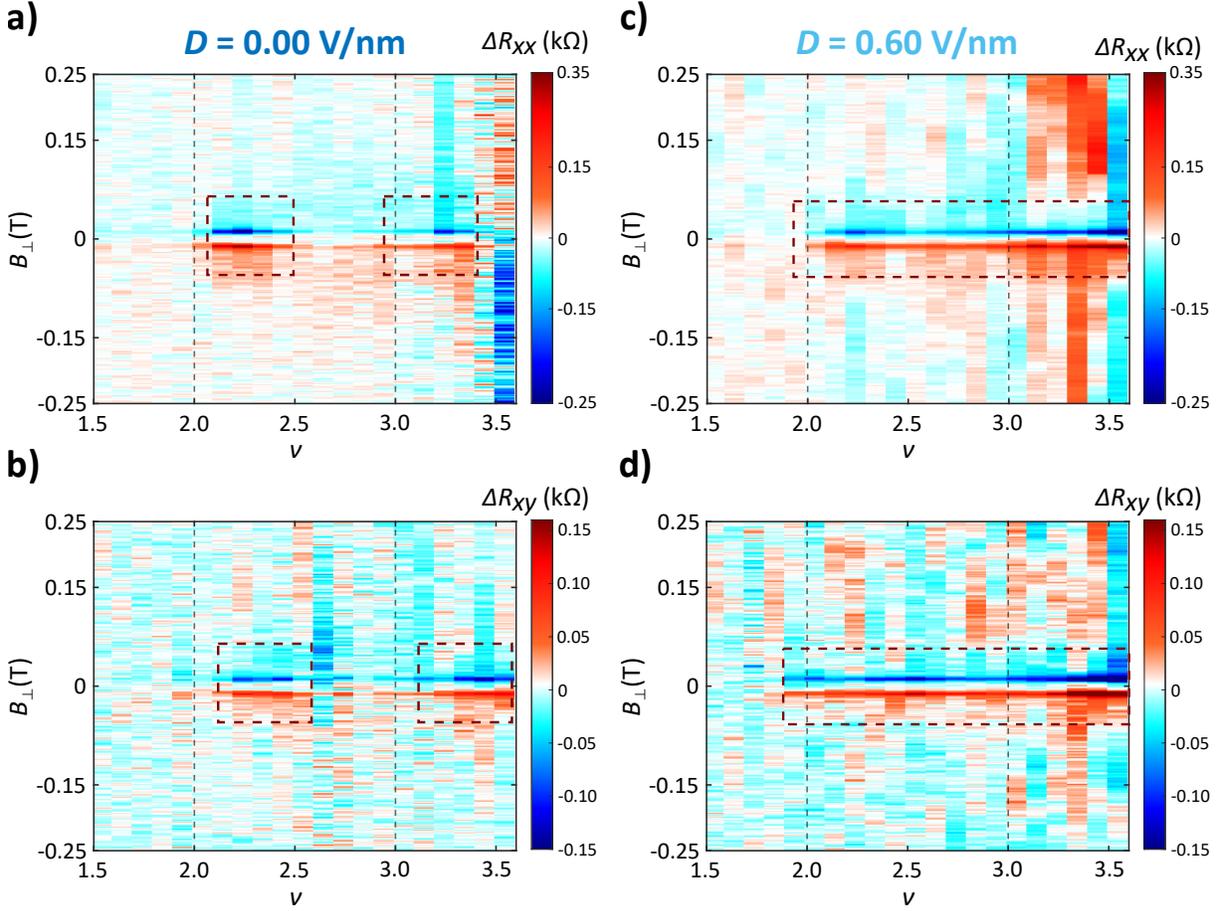

**SI-Fig. 26: Hysteresis in perpendicular magnetic field:** 2−d colormap of **a)** $\Delta R_{xx}(\nu, B_\perp)$ and **b)** $\Delta R_{xy}(\nu, B_\perp)$ for $D = 0.00$ V/nm at the effective sample/base temperature of 70 mK with a ramp rate of 100 mT/min. 2−d colormap of **c)** $\Delta R_{xx}(\nu, B_\perp)$ and **d)** $\Delta R_{xy}(\nu, B_\perp)$ for $D = 0.60$ V/nm with a ramp rate of 100 mT/min. We see that $\Delta R_{xx}$ for $B_\perp$ hysteresis 'islands' are situated analogously to the one seen for $B_{||}$ for both $\Delta R_{xx}$ and $\Delta R_{xy}$ centred around $\nu \sim 2.2$ and $\sim 3$ for $D = 0.00$ V/nm. These hysteresis islands is bridged together by a weaker emergent hysteresis phase at $D = 0.60$ V/nm, again in line with the phase observed for higher $D$ with $B_{||}$. It is important to mention here though, the magnitudes of $|\Delta R_{xx}|$ for $B_\perp$ are relatively quite small than for what we get for $B_{||}$.

## SI- 18   : Cernox sensor

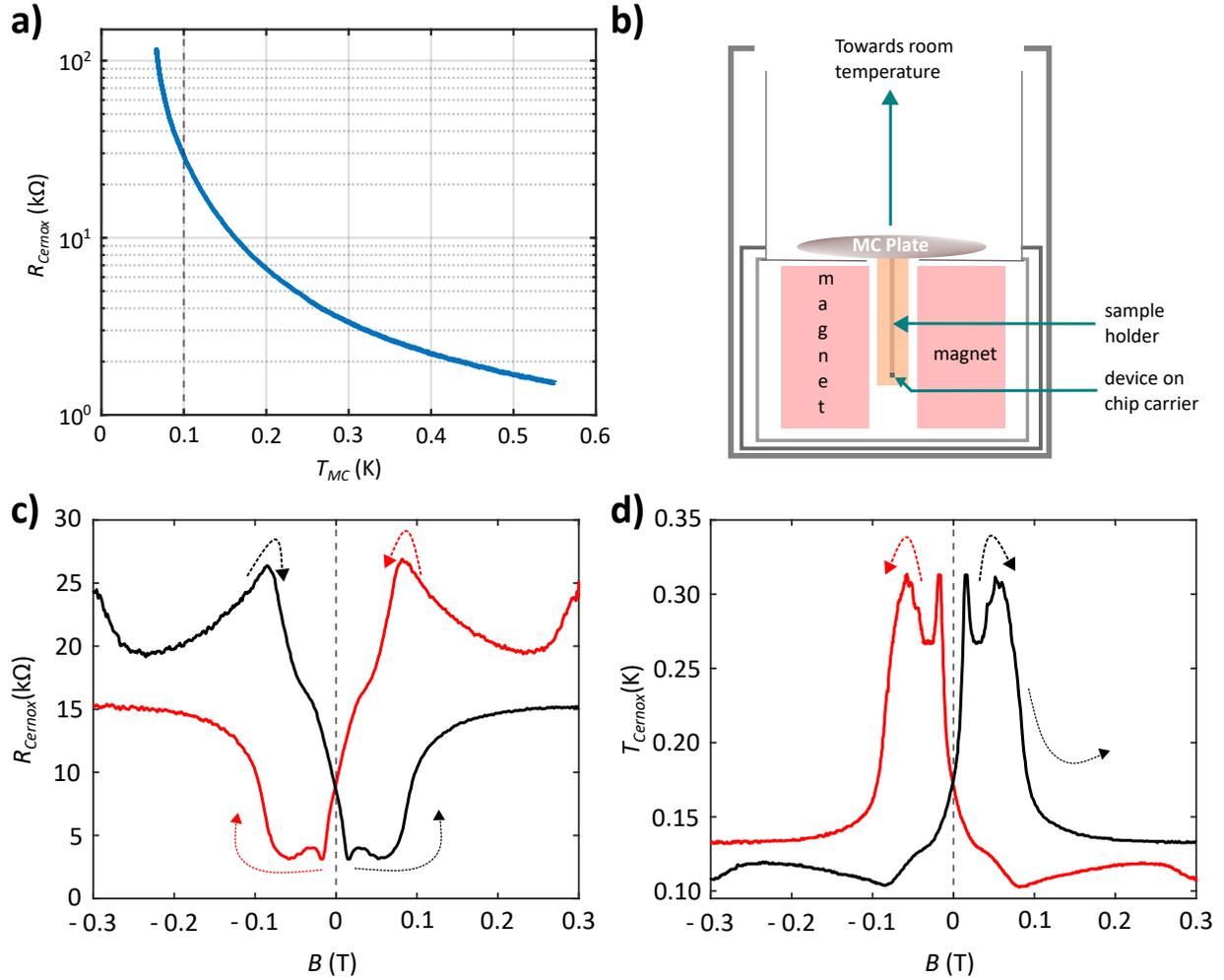

**SI-Fig. 27**: **Local sample stage temperature in $B$ sweep probed with Cernox sensor:** **a)** Cernox temperature $R$ vs. $T$ calibration curve. **b)** A schematic arrangement of the superconducting magnet, sample holder, and MC plate inside the dilution fridge. The plane of the sample site on the sample stage is nearly 2 feet away from the MC plate stage. So, the magnetic sweep-induced temperature change is much more prominent at the device plane compared to the MC plate stage. **c)** $R_{Cernox}$ and **d)** equivalent sample stage temperature $T_{Cernox}$. We observe $T_{Cernox}$ hysteresis profile analogous to the $R_{xx}$ hysteresis in TBG.



\end{document}